\documentclass[letterpaper,aps,prb,twocolumn,amsmath,showpacs,amssymb,10pt]{revtex4}
\pdfoutput=1
\usepackage{color}
\usepackage[pdftex]{graphicx,hyperref}
\usepackage{bm}
\usepackage{amsmath,amssymb}

\begin{document}

\title{Exact solution for square-wave grating covered with graphene: 
Surface plasmon-polaritons in the THz range}

\author{N. M. R. Peres$^1$, Yu. V. Bludov$^1$, Aires Ferreira$^2$, and M. I. Vasilevskiy$^1$ }

\affiliation{$^1$Department of Physics and Center of Physics,
 University of Minho, Campus de Gualtar, P-4710-057, Braga, Portugal}

\affiliation{$^2$Graphene Research Centre and Department of Physics, National University
of Singapore, 2 Science Drive 3, Singapore 117542}

\date{\today}

\begin{abstract}
We provide an analytical solution to the problem of scattering of electromagnetic radiation by
a square-wave grating with a flat graphene sheet on top. We show that for deep groves there is
a strong plasmonic response with light absorption in the graphene sheet reaching more than 45\%,
due to the excitation of surface plasmon-polaritons. The case of grating with a graphene sheet 
presenting an induced periodic modulation of the conductivity is also discussed.
\end{abstract}


\maketitle
\section{Introduction}

Plasmonic effects in graphene is currently an active research topic. The strong
plasmonic response of graphene at room temperature is tied up to its optical response, 
which can be controlled externally in different ways. The unique features of the optical conductivity
of single-layer graphene stem from the Dirac-like nature of quasi-particles, \cite{rmp,rmpPeres,rmpSarma}
and have been extensively studied in the past few years, both theoretically
\cite{nmrPRB06,falkovsky,stauberBZ,stauberphonons, StauberGeim,carbotte,Juan,Mishchenko,rmp,rmpPeres,LiLouie,PRL,aires,APLPhotonic,Platero} 
and experimentally,\cite{nair,kuzmenko,mak,basov,Crommieopt,kuzmenko2,kuzmenkoFaraday,NatureLoh}
including in the terahertz (THz) spectral range.\cite{MeasureTHzVis,AvourisIR,Ren,Ren2}

Indeed graphene holds many promises for cutting edge THz applications,\cite{THz}
which would be able to fill the so called {\it THz gap}.
More recently the interest has been focused to how graphene interacts with electromagnetic radiation in the THz.
\cite{avouris,koppens,koppensphoto,BasovPlasmonsI,BasovPlasmonsII,Pellegrini,KuzmenkoPlasmons}
One of the goals is to enhance the absorption of graphene for the development of more efficient photodetectors
in that spectral range. This can be done in several different ways, by 
(i) producing micro-disks of graphene on a layered structure;\cite{avouris}
(ii) exploiting the physics of quantum dots and metallic arrays on graphene;\cite{Echtermeyer,koppensphoto}
(iii) using a graphene based grating;\cite{nunoSPP,YuliyPRB,AiresPeres,Gao}
(iv) putting graphene inside an optical cavity;\cite{PeresCavity,MuellerCavity} and
(v) depositing graphene on a photonic crystal.\cite{APLPhotonic}
In cases (i), (ii), and (iii) the excitation of plasmons \cite{Dubinov,Xu,Kostia,Soukoulis} 
is responsible for the enhancement of the absorption. In case (iv)
photons undergo many round trips inside the cavity enhancing the chances
of being absorbed by graphene. In case (v) the authors consider a photonic crystal
made of SiO$_2$/Si. In the visible range of the spectrum the dielectric constants
of SiO$_2$ and Si differ by more than one order of magnitude and choosing the
width of the  SiO$_2$/Si appropriately it is possible to induce a large photonic
band gap in the visible range. Combining the presence of the band gap
with an initial spacer layer the absorption can be enhanced by a factor of four.
In the case studied in Ref. \onlinecite{APLPhotonic} the optical conductivity of
graphene is controlled by interband transitions. Although this work\cite{APLPhotonic} 
focused on the visible spectrum, there is {\it a priori} no reason why the same
principle cannot be extended to the THz.

The physics of surface plasmon-polaritons in graphene has also been explored
for the development of devices for optoelectronic applications.\cite{Ferrari,Loh,Modulators} Such
devices include  optical switches \cite{YuliyEPL} and polarizers.\cite{YuliyPolarizer}
It has been shown that metallic single-wall carbon nanotubes, which have a linear spectrum close to zero energy, can act as  polarizers as well.
\cite{Ren} Theoretical studies of the optical response of graphene under 
intense THz radiation has also been performed.\cite{Wu,Mikhailov1,Mikhailov2,Mikhailov3}

In general, the conductivity of graphene is a sum of two contributions: (i) a Drude--type term,
describing intraband processes and (ii) a term describing interband transitions. At zero
temperature the optical conductivity has a simple analytical expression.
\cite{nmrPRB06,falkovsky,Falkovsky2,rmp,rmpPeres,StauberGeim} In what concerns our study,
the physics of the system is dominated by the intraband contribution,\cite{FengWang}
which reads
\begin{equation}
\sigma_D =\sigma_0\frac{4E_F}{\pi}\frac{1}{\hbar\Gamma-i\hbar\omega}\,,
\label{eq_sigma_xx_semiclass}
\end{equation}
where, $\sigma_0=\pi e^2/(2h)$,
$\Gamma$ is the relaxation rate, $E_F>0$ is the
Fermi level position with respect to the Dirac point, and $\omega$ is the radiation
frequency. It should be noted that $\sigma_D$
has a strong frequency dependence and is responsible for the optical behaviour
of graphene in the THz spectral range.

\begin{figure}[ht]
 \begin{center}
 \includegraphics*[width=8cm]{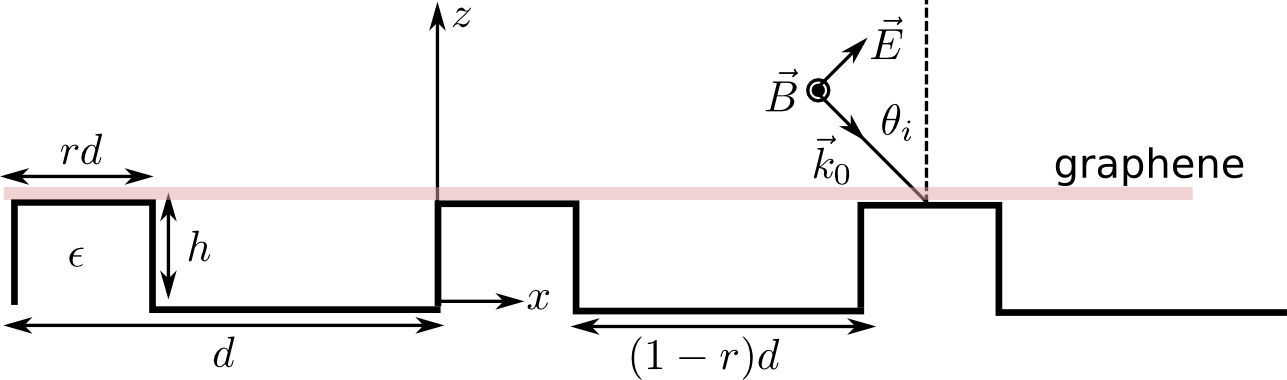}
 \end{center}
\caption{Geometry of the problem. The incident angle is $\theta_i$ and the 
incoming wave number is $\bm k_0$. The radiation is $p-$polarized. A top gate for the electrostatic doping of graphene can be arranged as a transparent electrode placed at some distance above the graphene sheet. Alternatively, a bottom gate can be placed below the dielectric substrate.}
\label{fig_square}
\end{figure}

The problem we consider in this work is
the scattering of electromagnetic radiation by graphene deposited on a grating with a 
square-wave profile, as illustrated in Fig.~\ref{fig_square}. We further assume that the radiation
is $p-$polarized (TM wave), that is, we take $\bm B=(0,B_y,0)$ and $\bm E=(E_x,0,E_z)$. 
The square-wave profile is rather pathologic due to the infinite derivative at the edges of the steps and therefore the usual methods fail.\cite{Chandezon,Lifeng,nunoSPP,AiresPeres} An alternative route is to obtain the exact eigenfunctions in the region of the grooves.\cite{Ping} Fortunately, the geometry of the problem is equivalent to that of a Kronig-Penney model appearing in the band theory of solids and for which an exact solution exists. We will show that Maxwell's equations can be put in a form equivalent to the Schr\"odinger equation for the Kronig-Penney model, and hence an exact solution for the fields in the region of the grooves is possible.

In this work we assume that graphene is doped. In a practical implementation, this can be achieved via gating or by chemical means.\cite{chemicaldopeII,chemicaldopeI} Remark that in a bottom gate structure the conductivity of graphene becomes position dependent along the $x-$direction.\cite{nunoSPP} In this case, the analysis presented below is incomplete (see, however, Sec.~\ref{sec_inho}, where a simplified model is analysed). Nevertheless, the results of Ref.~\onlinecite{nunoSPP} show that a position dependent conductivity alone already induces surface plasmon-polaritons on graphene. Therefore, a spatially modulated optical response, when combined with a grating, will enhance the excitation of surface plasmon-polaritons, as confirmed by the analysis given in Sec.~\ref{sec_inho}. For the top gate configuration and for chemical doping, the problem of a spatial dependent conductivity does not arise.\cite{LongJuPlasmonics} 

It can be argued that in the geometry we are considering (see Fig.~\ref{fig_square}) the portion of graphene over the grooves will be strained. Clearly, experimental setups can be designed as to overcome (at least partially) such effects e.g.,~by filling the grooves with a different dielectric and using a top gate or a chemically doped graphene. However, in the bottom gate configuration, even with filled grooves, graphene will have a  position dependent conductivity. A complete description should thus include all contributions: the grating effect itself, inhomogeneous doping, and the strain fields in graphene. In this work we assume that the graphene sheet remains flat. First, the optical conductivity will be taken homogeneous in order to study solely the effect of dielectric grating in the presence of uniform graphene. The additional effect of periodic modulation of the graphene conductivity, with the same period as the grating, will be considered in Sec.~\ref{sec_inho}. We believe that our calculations are accurate for the cases of either electrostatic doping by a top gate,\cite{LongJuPlasmonics} or doping by chemical means.\cite{chemicaldopeII,chemicaldopeI}

\section{Exact eigenmodes in the grating region}
\label{sec_solution}

For TM polarized light with angular frequency $\omega $ (see Fig.~\ref{fig_square}) the Helmoltz equation assumes the simple form 
\begin{equation}
 (\Delta_{\parallel}+\mu_0\epsilon\omega^2) B_y(x,z)=0\,,
 \label {Helm} 
 \end{equation}
where $\mu_0$ is the vacuum permeability and $\Delta_{\parallel}=\partial^2_x+\partial^2_z$. Since the boundary between the vacuum and the dielectric is piecewise, the Helmholtz equation  holds true in the regions of width $rd$ and $(1-r)d$ with the appropriate change of the dielectric constant, $\epsilon$ in the former and $\epsilon_0$ in the latter. For $z>h$ the dielectric constant is homogeneous and equal to $\epsilon_0$, whereas for $z<0$ it is $\epsilon=\epsilon_0\epsilon_r$, and $\epsilon_r$ is the relative permittivity.

We search solutions of Eq.~\ref{Helm} in the form
\begin{equation}
 B_y=B_y(x,z)=X(x)e^{\pm i\Lambda z}\,,
\end{equation}
where $\Lambda$ is a constant. With this ansatz, the Helmholtz equation reduces to 
\begin{equation}
 \partial_x^2X(x)=(\Lambda²-\mu_0\epsilon\omega^2)X(x)\,,
\end{equation}
whose solution is 
\begin{equation}
 X(x)=A_je^{ik_jx}+B_je^{-ik_jx}\,,
\end{equation}
and where $k_j^2$ is given by
\begin{equation}
 k_j^2=\mu_0\epsilon_j\omega^2-\Lambda^2,
 \label {k_j}
\end{equation}
with the subscript $j=1$ [$j=2$] referring to the region $(1-r)d$  [$rd$]. 
Putting all pieces together, the field component $B_y$ in the regions of the grooves reads
\begin{equation}
 B_y^{(j)}=[A_je^{ik_jx}+B_je^{-ik_jx}]e^{\pm i\Lambda z}\,.
\label{Byj}
\end{equation}
The determination of $\Lambda$ in Eq.~(\ref{Byj}) leads to an eigenvalue problem that will be considered in the next section. We note that $\Lambda$ can be either real or pure imaginary: reals values correspond to propagating diffracted orders, whereas imaginary ones correspond to evanescent waves. From Maxwell's equations it follows that the electric field components in the region $0<z<d$ read:
\begin{eqnarray}
 E_x^{(j)}&=&\frac{\pm \Lambda}{\mu_0\epsilon_j\omega}
[A_je^{ik_jx}+B_je^{-ik_jx}]e^{\pm i\Lambda z}\,,\\
 E_z^{(j)}&=&\frac{-k_j}{\mu_0\epsilon_j\omega}
[A_je^{ik_jx}-B_je^{-ik_jx}]e^{\pm i\Lambda z}\,.
\end{eqnarray}
As shown in what follows, there is a relation between the coefficients $A_j$ and $B_j$, which is obtained from the solution of the eigenvalue problem.

\section{Transfer matrix and the eigenvalue equation}

\begin{figure}[ht]
 \begin{center}
 \includegraphics*[width=8cm]{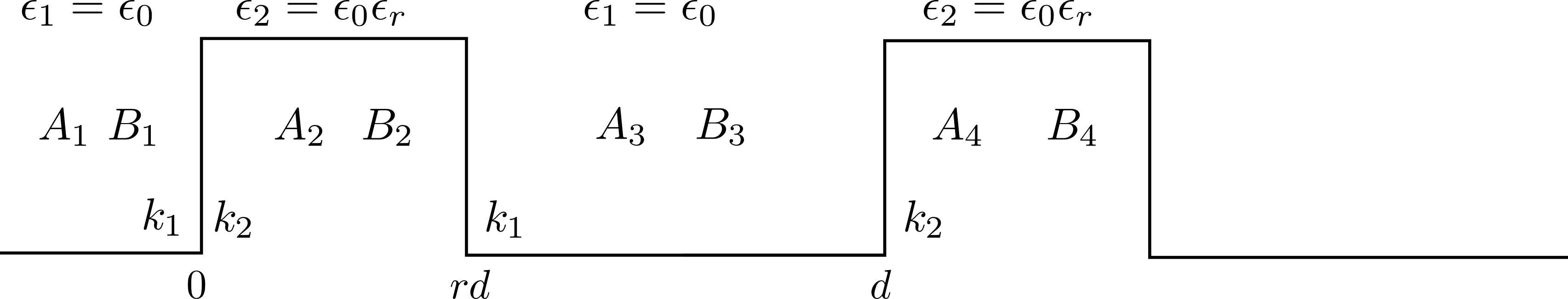}
 \end{center}
\caption{Effective stratified medium along the $x-$direction in the region $0<z<h$.}
\label{fig_stratified}
\end{figure}

Along the $x-$direction, and for $0<z<h$, we have a stratified medium with alternating dielectric constants. 
Then we can  relate
the amplitudes $A_j$ and $B_j$ with $A_{j+1}$ and $B_{j+1}$ using the 
boundary conditions at the interfaces of the two
dielectrics. Furthermore, using the Bloch theorem we can find an eigenvalue equation for the parameter
$\Lambda$. The amplitudes in the different regions of the stratified medium are represented in 
Fig.~\ref{fig_stratified}.
The continuity of the $z$-component of the electric field at the boundary $x=rd$ imposes a relation between the amplitudes $A$ and $B$ in regions $j=2,3$ (refer to Fig.~\ref{fig_stratified} for the definition of the different regions), i.e.,
\begin{equation}
\frac{k_2}{\epsilon_r}[A_2e^{ik_2rd}-B_2e^{-ik_2rd}]=k_1[A_3e^{ik_1rd}-B_3e^{-ik_1rd}]\,. 
\end{equation}
Note that according to the definition in Eq.~(\ref {k_j}), we must have $k_3=k_1$ and $k_4=k_2$. At the same boundary, the continuity of the magnetic field yields
\begin{equation}
A_2e^{ik_2rd}+B_2e^{-ik_2rd}=A_3e^{ik_1rd}+B_3e^{-ik_1rd}\,. 
\end{equation}
It is convenient to write these two equations in matrix form as follows
\begin{equation}
 \left[
\begin{array}{c}
 A_2\\
B_2
\end{array}
\right]=\Phi(k_2rd)K_2^{-1}K_1\Phi(-k_1rd)
 \left[
\begin{array}{c}
 A_3\\
B_3
\end{array}
\right]\,,
\label {T2-3}
\end{equation}
where
\begin{eqnarray}
 \Phi(x)&=&\left[
\begin{array}{cc}
e^{-ix} & 0 \\
0 & e^{ix}
\end{array}
\right]\,,\\
K_1&=&\left[
\begin{array}{cc}
k_1 & -k_1 \\
1 & 1
\end{array}
\right]\,,\\
K_2&=&\left[
\begin{array}{cc}
k_2/\epsilon_r & -k_2/\epsilon_r \\
1 & 1
\end{array}
\right]\,.
\end{eqnarray}
Similar continuity conditions apply to the boundary at $x=d$, resulting in the following constraints
\begin{equation}
k_1[A_3e^{ik_1d}-B_3e^{-ik_1d}]=\frac{k_2}{\epsilon_r}[A_4e^{ik_2d}-B_4e^{-ik_2d}]\,,
\end{equation}
\begin{equation}
A_3e^{ik_1d}+B_3e^{-ik_1d}=A_4e^{ik_2d}+B_4e^{-ik_2d}\,. 
\end{equation}
As before, these equations can be written in matrix form as
\begin{equation}
 \left[
\begin{array}{c}
 A_3\\
B_3
\end{array}
\right]=\Phi(k_1d)K_1^{-1}K_2\Phi(-k_2d)
 \left[
\begin{array}{c}
 A_4\\
B_4
\end{array}
\right]\,,
\label{constr2}
\end{equation}
Combining Eqs.~(\ref{T2-3}) and (\ref{constr2}), we arrive at the following result
\begin{eqnarray}
  \left[
\begin{array}{c}
 A_2\\
B_2
\end{array}
\right] &=&
\Phi(k_2rd)K_2^{-1}K_1\Phi[k_1d(1-r)]K_1^{-1}K_2
\nonumber\\
&\times&
\left[
\begin{array}{c}
 A_4\\
B_4
\end{array}
\right]\:,
\end{eqnarray}
where a global phase was absorbed in the coefficients $A_4$ and $B_4$.
The transfer matrix,
\begin{equation}
 \mbox{T}=\left [ \Phi(k_2rd)K_2^{-1}K_1\Phi[k_1d(1-r)]K_1^{-1}K_2\right ]^{-1}\:,
\end{equation}
propagates the field amplitudes through the 1D crystal in the $x-$direction,
\begin{eqnarray}
  \left[
\begin{array}{c}
 A_4\\
B_4
\end{array}
\right] &=&  \mbox{T}
\left[
\begin{array}{c}
 A_2\\
B_2
\end{array}
\right]\:.
\label {T2-4}
\end{eqnarray}
[Note that $ \mbox{det (T)}=1$.] On the other hand, by virtue of the Bloch theorem we have:
\begin{eqnarray}
  \left[
\begin{array}{c}
 A_4\\
B_4
\end{array}
\right] &=& e^{ik_0 d \sin \theta _i}
\left[
\begin{array}{c}
 A_2\\
B_2
\end{array}
\right]\:,
\label {Bloch}
\end{eqnarray}
where $k_0 \sin\theta_i$ is the $x-$component of the wavevector of the incoming electromagnetic radiation (Fig.~\ref{fig_square}). We thus arrive at the important intermediate result
\begin{equation}
e^{ik_0d\sin\theta_i} \left[
\begin{array}{c}
 A_2\\
B_2
\end{array}
\right] =\rm{T}
\left[
\begin{array}{c}
 A_2\\
B_2
\end{array}
\right]
\,.
\label{eq_Bloch_eq}
\end{equation}
The compatibility condition of Eq.~(\ref {eq_Bloch_eq}) provides the eigenvalue equation
\begin{equation}
2\cos(k_0d\sin\theta_i)=\mbox{Tr(T)}\,, 
\end{equation}
or explicitly,
\begin{eqnarray}
 &&2\cos(k_0d\sin\theta_i)=2\cos[k_1d(1-r)]\cos(k_2dr)
\nonumber\\
&-&
\frac{\epsilon_r^2k_1^2+k_2^2}{\epsilon_rk_1k_2}
\sin[k_1d(1-r)]\sin(k_2dr)\,.
\label{eq_eigenequation}
\end{eqnarray}
Eq. (\ref{eq_eigenequation}) allows for the determination of the permitted 
values of $\Lambda ^2$ and is very similar to the eigenvalue equation of the Kronig-Penney
model of electron bands. 

The relation (\ref {T2-3})
allows us to express the coefficients $A_3$ and $B_3$ in terms of $A_2$ and $B_2$.
Thus, the function $X(x)$ over the whole unit cell can be written in terms of 
$A_2$ and $B_2$ only. Furthermore, Eq.~(\ref{eq_Bloch_eq}) gives a relation between the coefficient $A_2$ and $B_2$,
\begin{equation}
 B_2 = \frac{-T_{11}+e^{ik_0d\sin\theta_i}}{T_{12}}A_2\,,
  \label{eq_B2_func_A2}
\end{equation}
where $T_{12}$ and $T_{22}$ are two matrix 
elements of the transfer matrix $ \mbox{T}$. Therefore, the function
$X(x)$ over the unit cell is proportional to the only coefficient, $A_2$. Note that the matrix elements contain $\Lambda ^2$, therefore we shall label the possible functions $X(x)$ by index $\ell $ running over all possible eigenvalues ($\pm \Lambda _\ell$).     
The function $X_\ell(x)$ in a unit cell has the form
\begin{equation}
X_\ell(x) = 
\left\{
\begin{array}{c}
 A_2e^{ik_2x} + B_2e^{-ik_2x}\,,\hspace{0.5cm} 0<x<rd\,,\\
 A_3e^{ik_1x} + B_3e^{-ik_1x}\,,\hspace{0.5cm} rd<x<d\,.
\end{array}
\right. 
\end{equation}
Equation~(\ref {T2-3}) can be writtent explicitly as
\begin{eqnarray}
\nonumber
 A_3 &=& aA_2+bB_2\,,\\
 \nonumber
 B_3 &=& c A_2+f B_2\,,
\end{eqnarray}
where 
\begin{eqnarray}
\nonumber
 a&=&\frac{1}{2k_1\epsilon_r}e^{-i(k_1-k_2)rd}(k_2+k_1\epsilon_r)\,,\\
\nonumber
 f&=&\frac{1}{2k_1\epsilon_r}e^{i(k_1-k_2)rd}(k_2+k_1\epsilon_r)\,,\\
\nonumber
 b&=&\frac{1}{2k_1\epsilon_r}e^{-i(k_1+k_2)rd}(-k_2+k_1\epsilon_r)\,,\\
\nonumber
 c&=&\frac{1}{2k_1\epsilon_r}e^{i(k_1+k_2)rd}(-k_2+k_1\epsilon_r)\,.
\end{eqnarray}
Since the wave numbers $k_j$ can be complex, $f$ is not necessarily the
complex conjugate of $a$; the same applies to $b$ and $c$. In terms of these coefficients, $X_\ell(x)$ reads:
\begin{equation}
X_\ell(x) = 
\left\{
\begin{array}{c}
 A_2e^{ik_2x} + B_2e^{-ik_2x}\,,\hspace{0.5cm} 0<x<rd\,,\\
(aA_2+bB_2)e^{ik_1x} +\\ 
(c A_2+f B_2)e^{-ik_1x}\,,\hspace{0.5cm} rd<x<d\,.
\end{array}
\right. 
\label{eq_coeffs}
\end{equation}
We also note that Eq. (\ref{eq_B2_func_A2}) allows to replace $B_2$ in Eq. (\ref{eq_coeffs}).
Finally, the magnetic field in the region $0<z<h$ (hereafter denoted as region II) has the form,
\begin{equation}
 B_y^{II}=\sum_{\ell} X_\ell(x)(C_\ell e^{i\Lambda_\ell z}+D_\ell e^{-i\Lambda_\ell z})\,,
 \label {BII}
\end{equation}
where the summation is over all the eigenvalues determined from the solution of 
Eq.~(\ref{eq_eigenequation}) and $C_\ell$ and $D_\ell$ are some coefficients that will be determined in the next section. With this we conclude the exact solution for the eigenmodes in the grating region.
\section{Solution of the scattering problem}

We now derive the equations for the scattering problem represented in Fig.~\ref{fig_square}.
For $z>h$ (region I) the magnetic field is written as 
\begin{equation}
 B^I_y=e^{i(\alpha_0x-\beta^{(1)}_0z)}
+\sum_{n=-\infty}^\infty R_ne^{i(\alpha_nx+\beta^{(1)}_nz)}\,,
\end{equation}
and for $z<0$ (region III) we have 
\begin{equation}
 B^{III}_y=\sum_{n=-\infty}^\infty T_ne^{i(\alpha_nx-\beta^{(2)}_nz)}\,,
\end{equation}
where
\begin{eqnarray}
 \alpha_n&=&k_0\sin\theta_i-2\pi n/d\,,\\
\beta^{(p)}_n&=&\left  \{
\begin{array}{c}
\sqrt{q^2_p-\alpha^2_n}\,; \hspace{0.5cm} q_p\ge \alpha_n\\
i\sqrt{\alpha^2_n-q^2_p}\,; \hspace{0.5cm} q_p<\alpha_n\
\end{array}
\right .\:,
\end{eqnarray}
with $q_p=\omega/v_p$ and the definition $q_1=k_0$; $n$ is an integer, $n\in[-\infty,\infty]$. Since we have four sets of unknown 
amplitudes, $C_\ell$, $D_\ell$, $R_n$, and $T_n$, and four boundary conditions,
two at $z=h$ and other two at $z=0$, we have a linear system of equations that can
be solved in closed form if truncated to some finite order, $N_\ell$, which is the number of the eigenvalues needed in Eq.~(\ref {BII})
for an accurate description of $B_y^{II}$  (we typically used $N_\ell\sim20$).
 
The boundary conditions at $z=0$ are
\begin{eqnarray}
 E_x^{III}(x,z=0) &=& E_x^{II}(x,z=0)\,,\\
 B_y^{III}(x,z=0) &=& B_y^{II}(x,z=0)\,,
\end{eqnarray}
whereas those at $z=h$ read
\begin{eqnarray}
 E_x^{I}(x,z=h) &=& E_x^{II}(x,z=h)\,,\\
 B_y^{I}(x,z=h) &-& B_y^{II}(x,z=h)= \nonumber\\
&-&\mu_0\sigma_D E_x^{II}(x,z=h)\,.
\end{eqnarray}
The latter represents the magnetic field discontinuity across the graphene sheet.\cite {YuliyEPL}  

These boundary conditions are $x$ dependent. Since the system has period $d$, we can eliminate
this dependence by multiplying the boundary conditions by $e^{-i\alpha_mx}$ and integrating over the unit cell. 
After some algebra, we arrive at 
\begin{eqnarray}
\label{eq_linear_1}
 \sum_{\ell=1}^{N_\ell}(f_{\ell m}^{(+)}C_\ell + f^{(-)}_{\ell m}D_\ell)&=&0\,,\\
\label{eq_linear_2}
\sum_{\ell=1}^{N_\ell} (g_{\ell m}^{(+)}C_\ell + g_{\ell m}^{(-)}D_\ell)&=&
2e^{-i\beta^{(1)}_0 h}\delta_{m,0}\,,
\end{eqnarray}
where
\begin{eqnarray}
\nonumber
 f^{(\pm)}_{\ell m}&=&\frac{\beta^{(2)}_m}{\epsilon}\chi_{\ell m}\pm\Omega_{\ell m}
\Lambda_\ell\,,\\
\nonumber
 g^{(\pm)}_{\ell m}&=&
\left(
\chi_{\ell m}\mp \frac{\sigma_D}{\omega}\Omega_{\ell m}\Lambda_\ell\mp
\epsilon_0\Omega_{\ell m}\frac{\Lambda_\ell}{\beta^{(1)}_m}
\right)
e^{\pm i\Lambda_\ell h}\:,
\end{eqnarray}
with 
\begin{eqnarray}
\label{eq_chilm}
 \chi_{\ell m}&=&\frac{1}{d}\int_0^d dx X_\ell(x)e^{-i\alpha_m x}\,,\\
\Omega_{\ell m}&=&\frac{1}{d}\int_0^d dx\frac{X_\ell(x)}{\epsilon(x)}e^{-i\alpha_m x}\,.
\end{eqnarray}
Here $\epsilon(x)$ is defined as $\epsilon$ [$\epsilon_0$] for $x=rd$ $[d(1-r)]$.  
Eqs.~(\ref{eq_linear_1}) and (\ref{eq_linear_2}) form a linear system of equations for the
amplitudes $C_\ell$ and $D_\ell$, from which the reflectance and transmittance
can be computed. Taking $N_\ell$ odd, the integer $m$ belongs to the 
interval $m\in[-(N_\ell-1)/2,(N_\ell-1)/2]$.
The transmittance and the reflectance amplitudes are given by
\begin{eqnarray}
 T_m&=& \sum_{\ell=1}^{N_\ell}\chi_{\ell m}(C_\ell + D_\ell)\,,\\
R_m &=& \delta_{0,m}e^{-2i\beta^{(1)}_0h}  + e^{-i\beta^{(1)}_mh} 
\nonumber\\
&\times&  
\sum_{\ell=1}^{N_\ell}\epsilon_0\Omega_{\ell m}\frac{\Lambda_\ell}{\beta^{(1)}_m}
(C_\ell e^{i\Lambda_\ell h}-D_\ell e^{-i\Lambda_\ell h})\,.
\end{eqnarray}
Since $X_\ell(x)$ is a sum of exponentials, the functions 
$\chi_{\ell m}$ and $\Omega_{\ell m}$ can be determined in closed form, which 
saves computational power. Explicit equations for $\chi_{\ell m}$ and 
$\Omega_{\ell m}$ are given in the Appendix \ref{ap_chi}.

\section{Results for homogeneous graphene}

\begin{figure}[ht]
 \begin{center}
 \includegraphics*[width=8cm]{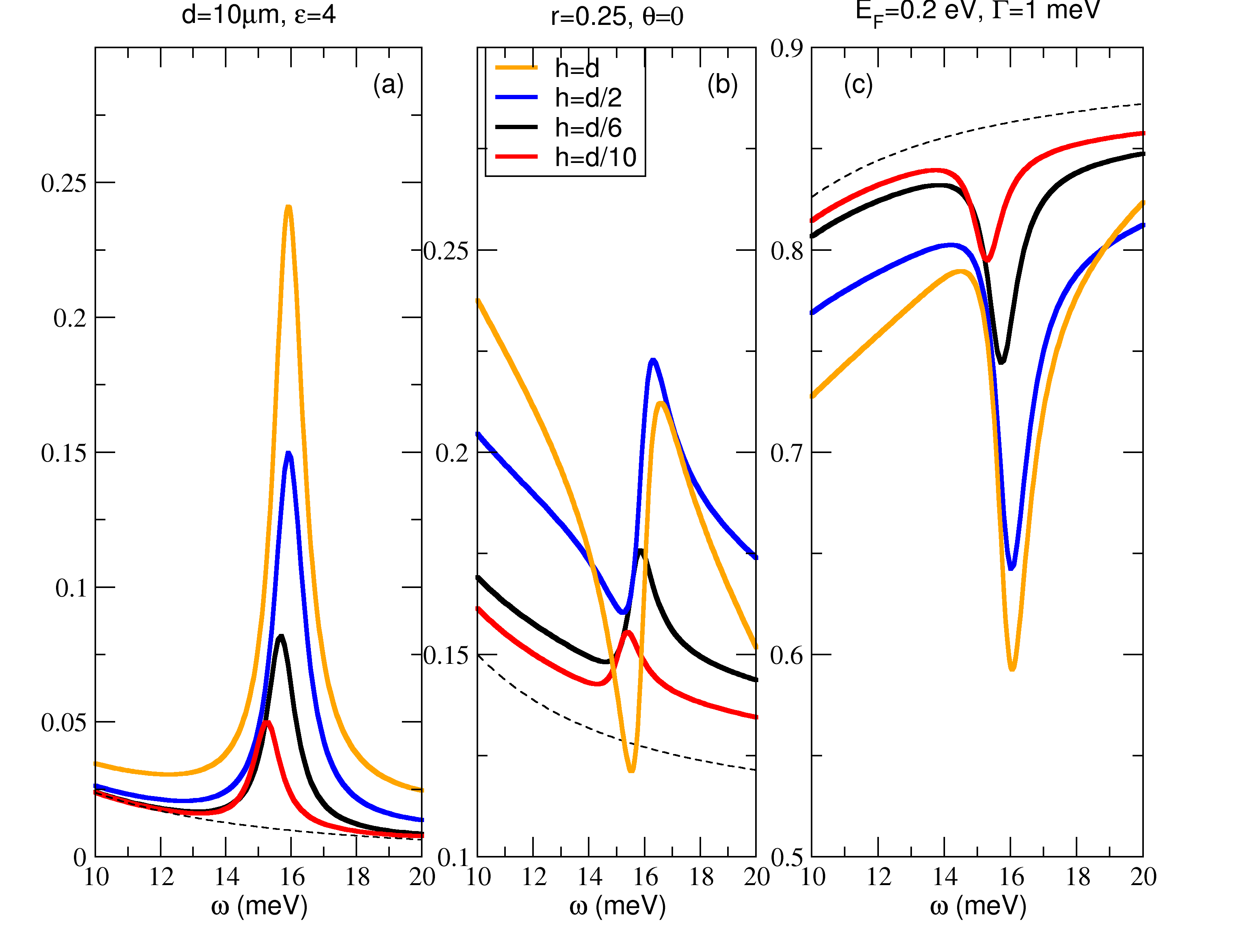}
 \end{center}
\caption{ Absorbance (a), reflectance (b), and transmittance (c) at normal incidence.
The parameters are $E_F=0.2$~eV, d=10~$\mu$m, $r=0.25$, $\theta _i=0$, $\epsilon_r=4$
(corresponding to SiO$_2$), and 
$\Gamma=1$~meV. The thin dashed line is the case $r=1$.
For reference, we note that 1 THz corresponds to an energy of 4.1~meV.}
\label{fig_absorbance}
\end{figure}

We now provide a number of results obtained from the solution of Eqs.~(\ref{eq_linear_1})
and (\ref{eq_linear_2}). In Fig.~\ref{fig_absorbance} we depict the absorbance (a), reflectance (b), and 
transmittance (c) at normal incidence. For the parameters considered only the specular
order exists; all the other diffraction orders are evanescent. Then, the absorbance
is defined as
\begin{equation}
{\cal A}=1-\vert R_0\vert^2-\frac{\beta_0^{(2)}}{\epsilon_r\beta_0^{(1)}}\vert T_0\vert^2\,. 
\end{equation}
A resonance is  clearly
seen in the absorbance curves at a frequency of about 16 meV, corresponding to the excitation of a 
surface plasmon-polariton. The position of the resonance depends on 
several parameters, given in the caption of Fig. \ref{fig_absorbance} and chosen as
typical values appropriate for THz physics.
The thin dashed curve corresponds to graphene on a homogeneous dielectric (no grating, $r=1$).
The effect of increasing the depth of the grooves, from $h=d/10$ up to 
$h=d$ is to produce an enhancement of the absorption, which for 
$h=d$ is almost of $25\%$ (in this case we are considering $r=0.25$).

The dispersion of a surface plasmon-polaritons in graphene,
when the sheet is sandwiched between two semi-infinite dielectrics, is,
in the  electrostatic limit, given by\cite{nunoSPP}   
\begin{equation}
\hbar\omega=\sqrt{2\alpha_f^{(\bar\epsilon)}E_Fc\hbar q}\,,
\label{eq_W_Plasmon_graphene}
\end{equation}
where
\begin{equation}
 \alpha_f^{(\bar\epsilon)} = \frac{e^2}{4\pi\bar\epsilon\hbar c}\,,
\label{eq_alpha}
\end{equation}
and $\bar\epsilon=\epsilon_0(1+\epsilon_r)/2$. Taking $q=2\pi/d$, the 
smallest lattice wavevector,
Eq. (\ref{eq_W_Plasmon_graphene})
predicts, for the parameters used in our calculation, a plasmon-polariton energy of 
\begin{equation}
 \hbar\omega = 12.1~\mbox{meV}\,,
\end{equation}
which is in the ballpark of the numeric result. We should however note that the position of 
the resonance depends
also on the parameter $r$, which is not captured by the above formula for the dispersion. 
We can include the effect of the parameter $r$ using an interpolative formula.
We define a new $\bar\epsilon$ as
\begin{equation}
 \bar\epsilon = \frac{\epsilon_0}{2}(2-r+\epsilon_rr)\,.
\end{equation}
Using this formula we obtain
\begin{equation}
 \hbar\omega =  16.3\mbox{ meV}\,,
\end{equation}
a much better approximation to the numerical value (15.9 meV).
To obtain the exact frequency of the resonance
we have to compute the plasmonic band structure due to the periodic dielectric,
that is, the band structure of a polaritonic crystal.\cite{YuliyPRB}
In the Appendix \ref{ap_polaritonic} we give the polaritonic band structure
of graphene on a square-wave grating.

\begin{figure}[ht]
 \begin{center}
 \includegraphics*[width=8cm]{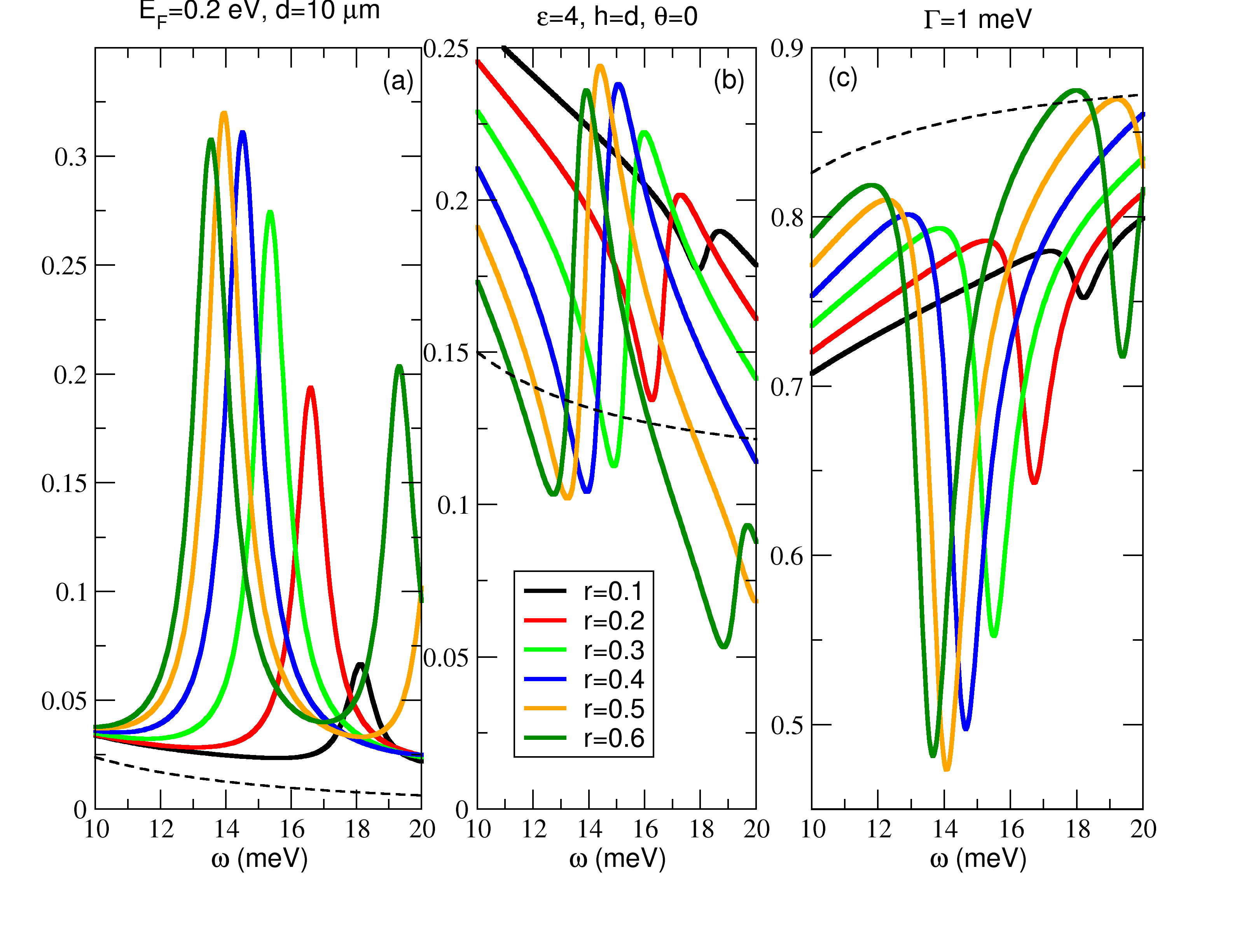}
 \end{center}
\caption{Absorbance (a), reflectance (b), and transmittance (c) at normal incidence.
The parameters are $E_F=0.2$~eV, d=10~$\mu$m, $h=d$, $\theta _i=0$, $\epsilon_r=4$, and 
$\Gamma=1$ meV. The thin dashed line is the case $r=1$.}
\label{fig_absorbance_function_of_r}
\end{figure}

In Fig.~\ref{fig_absorbance_function_of_r} we depict the 
absorbance (a), reflectance (b), and transmittance (c), as function of frequency,
for different values of $r$, from $r=0.1$ up to $r=0.6$, and 
keeping $d=h$, that is the limit of deep grooves.
As $r$ increases the 
position of the resonance shifts to the left. This happens because the 
width of the dielectric underneath graphene is increasing with $r$. Then,
according to Eq.~(\ref{eq_alpha}), the effective dielectric constant of the 
system increases, following from Eq.~(\ref{eq_W_Plasmon_graphene}) that
the resonance shifts toward lower energies.
In the case of $r=0.6$ {\it two} resonances are seen in the frequency window considered.
They correspond to the excitation of surface plasmon-polaritons of wave numbers
$2\pi/d$ and $4\pi/d$. Then, according to Eq.~(\ref{eq_W_Plasmon_graphene}),
the position of the second resonance should be $\sqrt 2$ times the first resonance frequency.
From the figure, the energy of the first resonance is $\hbar\omega=13.6$ meV, while that of the second one is 
$\hbar\omega=19.3$ meV; now we note that $\sqrt 2\times 13.6 = 19.2$ meV, in agreement
with the numerical result. We also note that the absorption of graphene is largest
when $r=0.5$, the symmetric case ($1-d=0.5$), reaching a value higher than $30\%$.
Although we do not show it here, we have found that the absorption 
grows monotonically with increasing  $h$ and attains a maximum at a value larger than 45\% for $h=2d$.

Finally, we note that the reflectance curves have Fano-type shape.\cite{YuliyPRB} This is due to the coupling
of the external radiation field with the excitation of Bragg modes of surface plasmon-polaritons
in periodically modulated structures. 

\begin{figure}
\includegraphics*[width=8.5cm]{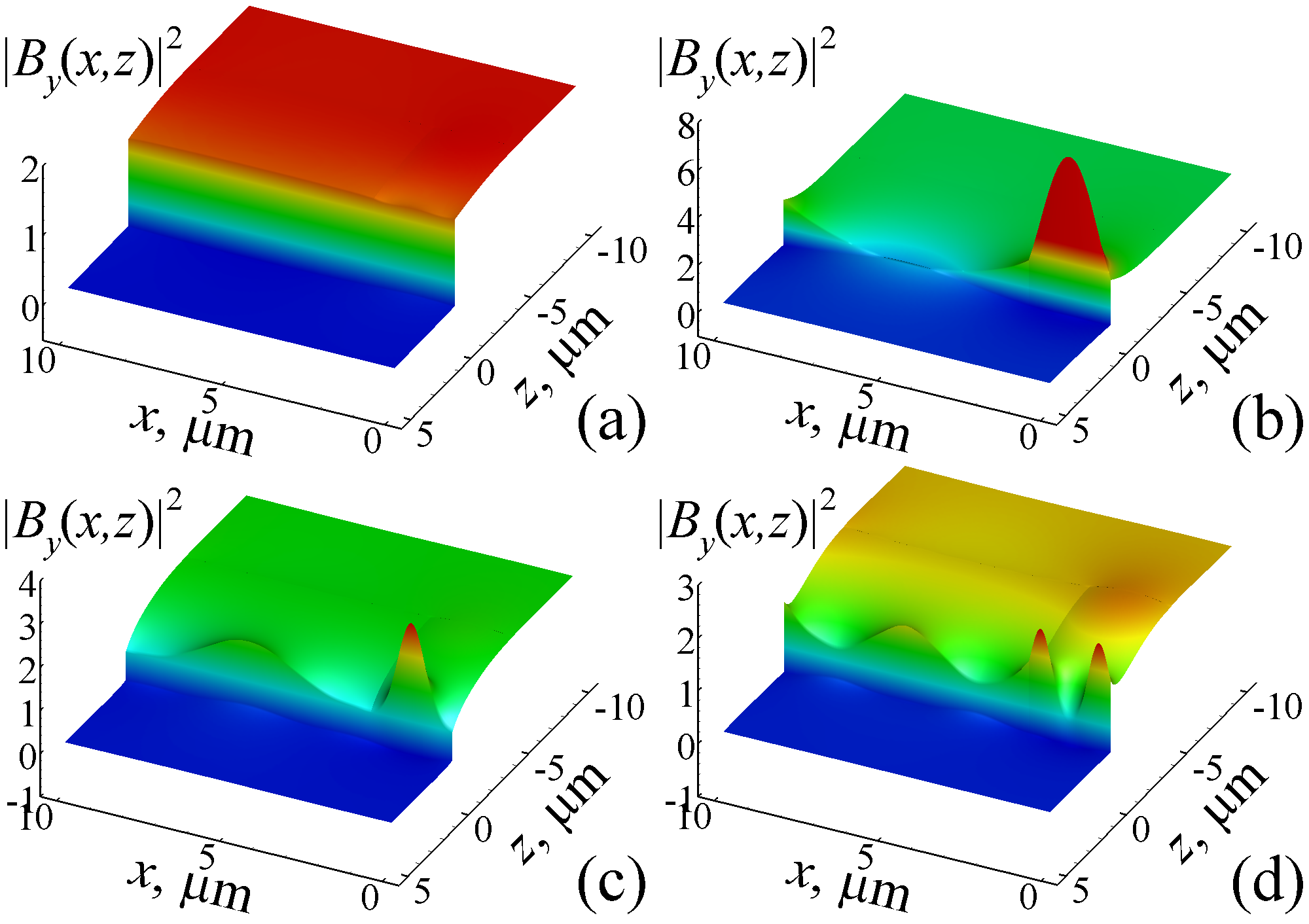}
\caption{Spatial dependence of the magnetic field $\left|B_{y}(x,z)\right|^{2}$,
diffracted from the graphene layer on top of the square-wave grating.  
We have considered a
groove depth $h=d/2$ and  different frequencies of
the incident wave: $\hbar\omega=13\,$meV (a), $\hbar\omega=16\,$meV (b), $\hbar\omega=23\,$meV
(c) and $\hbar\omega=28.4\,$meV (d). The other parameters of the structure
are the same as in Fig. \ref{fig_absorbance} }
\label{fig:fields}
\end{figure}

An important quantity when dealing with plasmonic effects is the enhancement of the electromagnetic
field close to the interface of the metal (in this case graphene) and the dielectric (the square-wave grating).
The spatial intensity of the diffracted electromagnetic field from
graphene is depicted in Fig.\ref{fig:fields}. For the case off-resonance
 {[}Fig.\ref{fig:fields}(a){]} the magnetic field has 
a relatively small amplitude 
and is  almost homogeneous along $x$-axis, while along $z$-axis
exhibits a discontinuity at the plane $z=0$ occupied by the graphene layer, in accordance
with the boundary conditions.

The situation changes dramatically when the incident wave frequency
coincides with that of a surface plasmon-polariton resonance {[}Figs.\ref{fig:fields}(b--d){]}.
In this case the amplitude of the electromagnetic field in the vicinity
of the graphene layer drastically increases. The electromagnetic field
amplitude is maximal when the surface plasmon-polariton resonance
occurs for  the first harmonic {[}Fig.\ref{fig:fields}(b){]}, that is for $q=2\pi/d$, while it
gradually decreases as the  harmonic's order increases (compare
with Fig.\ref{fig:fields}(c) for second and Fig.\ref{fig:fields}(d)
for third harmonics, respectively). It is this enhancement of the electromagnetic
field due to the excitation of surface plamons-polaritons in the vicinity of 
a metallic interface that is at the heart of sensing devices.

\section{Modulated doping: a simple model}
\label{sec_inho}

As discussed in the Introduction, doping graphene using a bottom gate
when the material lies on a grating
leads to a position dependent conductivity, which is periodic along the 
$x-$direction. In this section we consider a simple model where, in addition
to the grating itself, the conductivity is position dependent and periodic.
In our model the conductivity is given by\cite{nunoSPP} 
\begin{equation}
 \sigma(x)=\sigma_D(1-\kappa\sin\frac{2\pi x}{d})\,,
\label{eq_profile}
\end{equation}
where $\kappa$ is a parameter controlling the degree of inhomogeneity. 
We also take $r=0.5$ in our figures. This particular choice of $r$ and $\sigma(x)$ makes
the latter commensurate with the profile of the grating. From the point of view of the calculation
we have to replace $\sigma_D$ by $\sigma(x)$ in the formulas given above. The conductivity is then
expanded in Fourier series as
\begin{equation}
\sigma(x)=\sum_{p=-\infty}^\infty\sigma_pe^{i2\pi xp/d}\,, 
\end{equation}
where 
\begin{equation}
 \sigma_p=\frac{1}{d}\int_0^ddx\sigma(x)e^{-i2\pi p x/d}\,.
\end{equation}
For the case of the profile defined in Eq.~(\ref{eq_profile}) the 
Fourier series reduces to three terms only, those referring to
$p=0,\pm1$. A non-sinusoidal profile for $\sigma(x)$ will have
more harmonics than just these three. Therefore, our model
can also be considered the first term in the Fourier expansion 
of a more complex profile for $\sigma(x)$.

Relatively to the previous case of homogeneous conductivity, 
what changes in the equations is the form of the function $g^{(\pm)}_{\ell m}$,
which now reads
\begin{equation}
 g^{(\pm)}_{\ell m}=
\left(
\chi_{\ell m}\mp \frac{\sigma_D}{\omega}h_{\ell m}\Lambda_\ell\mp
\epsilon_0\Omega_{\ell m}\frac{\Lambda_\ell}{\beta^{(1)}_m}
\right)
e^{\pm i\Lambda_\ell h}\,,
\end{equation}
where 
\begin{equation}
h_{\ell m} =  \Omega_{\ell m} -\frac{\kappa}{2i}\Omega_{\ell m+1}
+\frac{\kappa}{2i}\Omega_{\ell m-1}\,.
\end{equation}
\begin{figure}[ht]
 \begin{center}
 \includegraphics*[width=8cm]{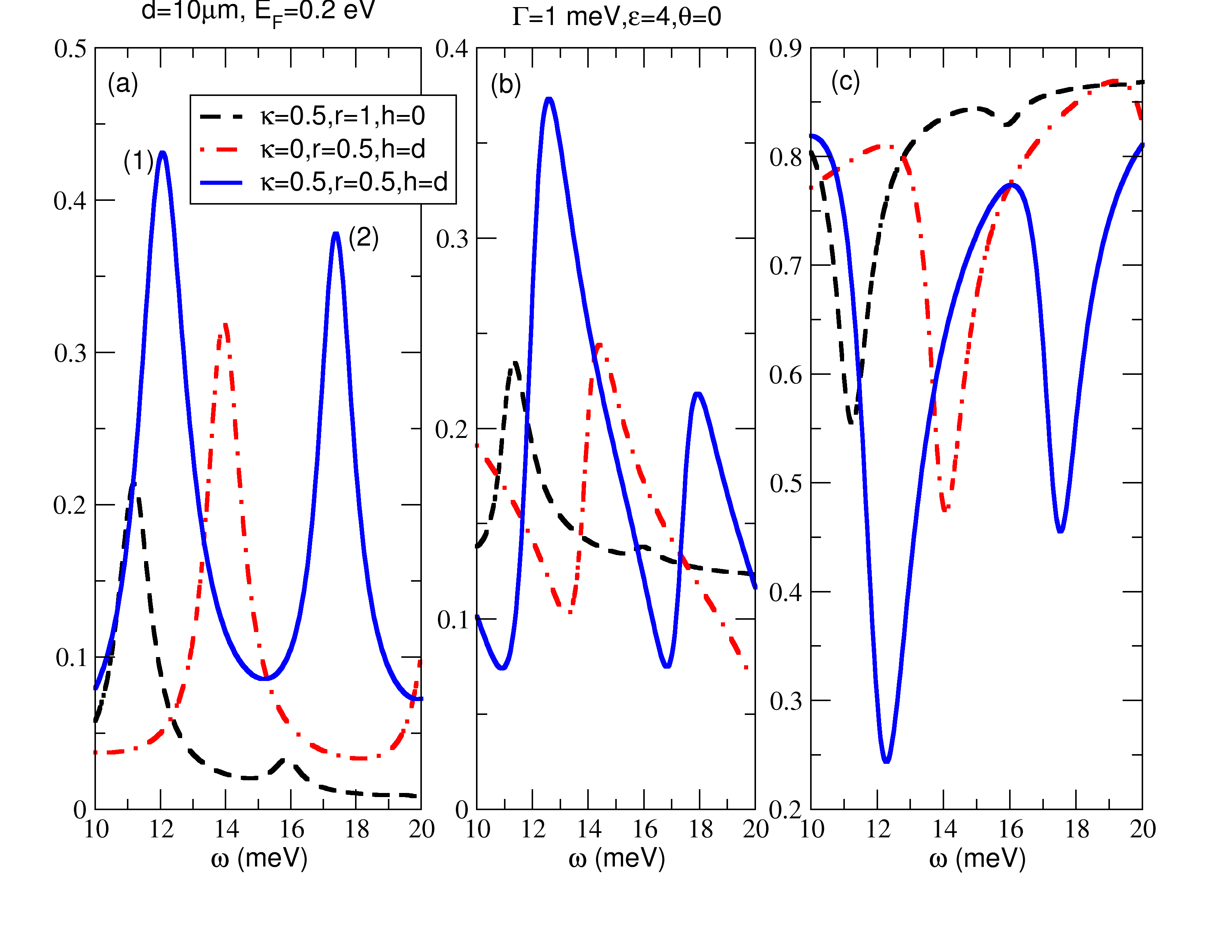}
 \end{center}
\caption{Absorbance (a), reflectance (b), and transmittance (c) at normal incidence.
Three distinct situations are considered: inhomogeneous conductivity, 
with grating (solid line) and without grating (dashed line),
and homogeneous conductivity with grating (dotted-dashed line). 
Other parameters as in Fig.~\ref{fig_absorbance_function_of_r}.}
\label{fig_sigma_of_x}
\end{figure}
The results for the absorbance, reflectance, and transmittance are given
in Fig.~\ref{fig_sigma_of_x}, considering the case where $\kappa=0.5$,
a fairly large value, corresponding to deep groves. 
The dashed line is the case where graphene has
an inhomogeneous conductivity on top of a homogeneous dielectric (i.e.,~no grating).
(This situation is artificial and is only included for the sake of
comparison.) Two resonant peaks can be seen, with the second peak 
being much smaller than the one at lower energies. The dashed-dotted line
is the case where we have homogeneous graphene on the grating,
and the solid line corresponds to the case where we have graphene with an inhomogeneous conductivity
on the grating. The main 
effect is a shift of the position of the resonant 
peaks toward lower energies and an enhancement of the 
absorption, which bears its origin on the combined effect of the grating and of
the periodic modulation of the conductivity. In the absorbance panel
the energy of the resonance labelled (2) is $\sqrt 2$ larger than that of the
resonance (1). Thus, the first resonance corresponds to an excitation
of a surface plasmon-polariton of wave number $q=2\pi/d$ whereas
the second peak corresponds to the excitation of  
surface plasmon-polariton of wave number $q=4\pi/d$. When we compare 
the dashed curve to the solid one we also note the prominence
of the second plasmonic peak in the latter case, which is almost as 
intense as the low energy one.  
This is a consequence of the square profile and the dielectric nature of the grating.
In metallic gratings the permittivity has a strong frequency dependence favouring the
the occurence of the most intense plasmonic resonances at lower energies.\cite{AiresPeres}

\section{Conclusions and future work}

We have shown that a flat sheet of graphene on top of a square-wave grating 
exhibits strong plasmonic behavior. With the help of the grating
one can create a surface plasmon-polariton resonance.
The effect is more pronounced in the case of
deep grooves, where absorbances higher than $45\%$ are attainable.
We note that the parameter $d$ is easy to control experimentally, providing a convenient way of tuning the position of the plasmonic resonance.
 The reflectance
curves show a Fano-type line shape, which is manifest of the 
coupling of the external electromagnetic field (with a continuum of modes) to the surface plasmons
in graphene (occupying a relatively narrow spectral band). The calculations we have performed assume that graphene is 
homogeneously doped (excluding Sec.~\ref{sec_inho}), 
 which is a  crude approximation for the case of a
 bottom gate configuration of graphene doping.
Therefore, our results are only directly applicable to the cases of either chemical
doping or doping by a top gate. This type of gating is within the state-of-the-art.\cite{LongJuPlasmonics}
In the particular case of bottom gate, we have to consider both the effect of 
inhomogeneous doping and the effect of strain (if the grooves are not filled
with another dielectric). Both effects result in a position 
dependent conductivity. Then, the calculation of the properties of surface plasmon-polaritons
requires the evaluation of the doping profile and the strain field. Once these are determined, one can use a Fourier expansion of the graphene conductivity as we have considered
in our phenomenological model. As these preliminary results show, the coupling of the external wave to the surface plasmon-polaritons can be significantly enhanced in the presence of both the dielectric grating and periodic modulation of the conductivity. Detailed calculations for realistic graphene conductivity profiles are the goal of a future work.

\section*{Acknowledgements}

NMRP acknowledges Bao Qiaoliang, Jos\'e Viana-Gomes, and Jo\~ao Pedro Alpuim for fruitful discussions. 
A. F. was supported by the National Research Foundation--Competitive Research Programme award ``Novel
2D materials with tailored properties: beyond graphene'' (Grant No. R-144-000-295-281). 
This work was partially supported by the Portuguese Foundation for Science and Technology (FCT) through 
Projects PEst-C/FIS/UI0607/2011 and PTDC-FIS-113199-2009.

\appendix
\section{Explicit form of $\chi_{\ell m}$}
\label{ap_chi}

The calculation of the integral in Eq. (\ref{eq_chilm}) has to be divided into two pieces. 
The function $\chi_{\ell m}$ can be written as
\begin{equation}
\chi_{\ell m}=\chi_{\ell m}^{(1)}+\chi_{\ell m}^{(2)}\,, 
\end{equation}
where 
\begin{eqnarray}
\chi_{\ell m}^{(1)}&=&-i\frac{A_2a}{d(k_1-\alpha_m)}(e^{i(k_1-\alpha_m)d}-e^{i(k_1-\alpha_m)rd})\nonumber\\
&+&  i\frac{A_2c}{d(k_1+\alpha_m)}(e^{-i(k_1+\alpha_m)d}-e^{-i(k_1+\alpha_m)rd})\nonumber\\
&-&  i\frac{B_2b}{d(k_1-\alpha_m)}(e^{i(k_1-\alpha_m)d}-e^{i(k_1-\alpha_m)rd})\nonumber\\
&+&  i\frac{B_2f}{d(k_1+\alpha_m)}(e^{-i(k_1+\alpha_m)d}-e^{-i(k_1+\alpha_m)rd})\,,
\end{eqnarray}
and
\begin{eqnarray}
\chi_{\ell m}^{(2)}&=&-i\frac{A_2}{d(k_2-\alpha_m)}(e^{i(k_2-\alpha_m)dr}-1)\nonumber\\
&+&  i\frac{B_2}{d(k_2+\alpha_m)}(e^{-i(k_2+\alpha_m)dr}-1)\,.
\end{eqnarray}
Since the square profile has a piecewise structure, the function $\Omega_{\ell m}$ is obtained
directly from the function $\chi_{\ell m}$ as
\begin{equation}
 \Omega_{\ell m} = \frac{\chi_{\ell m}^{(1)}}{\epsilon_0} + \frac{\chi_{\ell m}^{(2)}}{\epsilon_r\epsilon_0}\,. 
\end{equation}
\section{Polaritonic spectrum}
\label{ap_polaritonic}

\begin{figure}[!b]
\includegraphics*[width=8cm]{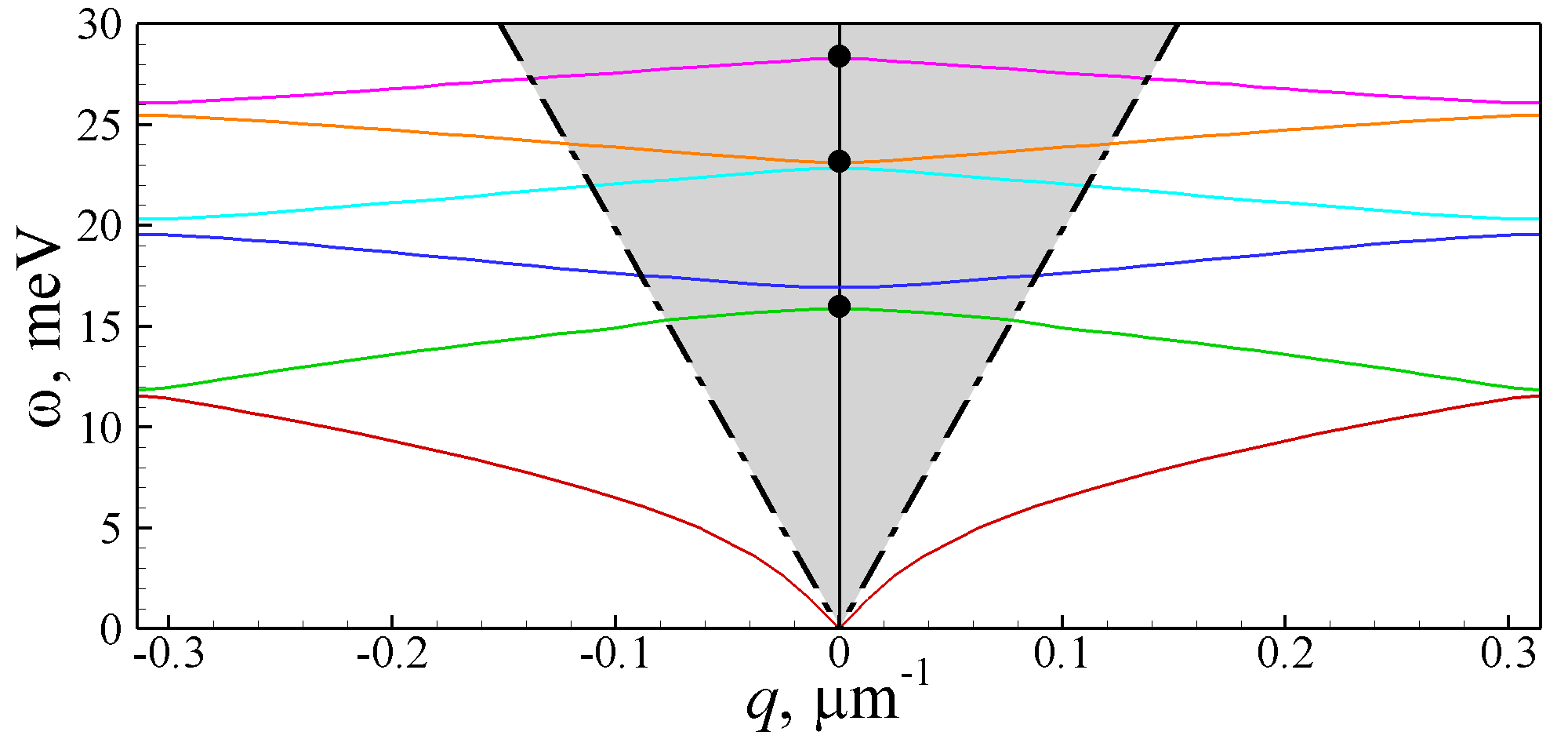}
\caption{Spectrum of surface plasmons-polaritons in graphene on a square-wave grating. 
We have considered $E_{F}=0.2\,$eV and 
$\Gamma=0$. The other parameters
are: $d=10\,\mu$m, $r=0.25$, $\epsilon_{r}=4$. The gray 
region represents the ``light cone'', limited by the light lines
$\omega=c|q|$ (dash-doted lines), inside which the excitation of
surface plasmons-polaritons by incident light is possible. The frequencies,
corresponding to the numerically obtained resonant minima of the transmittance
for normal incidence, $\theta_i=0$, are represented by black circles on
the line $q=0$.}
\label{fig:spectrum}
\end{figure}

The polaritonic spectrum, $\hbar\omega(q)$, of the surface plasmons-polaritons (SPP)
in graphene
on a square-wave grating 
is represented in Fig.\ref{fig:spectrum}, where $q=k_{0}\sin\theta_{i}$
is the Bloch wavenumber. Here, and in order to avoid the appearance of an
imaginary part of the eigenvalues, corresponding to surface plasmon-polariton
damping, we considered graphene without disorder ($\Gamma=0$). As
 expected, the periodicity of the grating induces 
a band structure in the SPP spectrum of graphene, showing energy gaps. At the same time, for  normal
incidence ($q=0$) the frequency of the second band ($\hbar\omega\approx15.85\,$meV)
almost  coincides with the numerically obtained resonant frequency
$\hbar\omega\approx15.9\,$meV. A similar good agreement happens for
the frequency of fifth band and that of second resonance ($\hbar\omega\approx23.1\,$meV
and $\hbar\omega\approx23\,$meV, respectively), as well as for frequency
of sixth band and that of third resonance ($\hbar\omega\approx28.3\,$meV
and $\hbar\omega\approx28.4\,$meV, respectively). 

\begin{thebibliography}{64}
\expandafter\ifx\csname natexlab\endcsname\relax\def\natexlab#1{#1}\fi
\expandafter\ifx\csname bibnamefont\endcsname\relax
  \def\bibnamefont#1{#1}\fi
\expandafter\ifx\csname bibfnamefont\endcsname\relax
  \def\bibfnamefont#1{#1}\fi
\expandafter\ifx\csname citenamefont\endcsname\relax
  \def\citenamefont#1{#1}\fi
\expandafter\ifx\csname url\endcsname\relax
  \def\url#1{\texttt{#1}}\fi
\expandafter\ifx\csname urlprefix\endcsname\relax\def\urlprefix{URL }\fi
\providecommand{\bibinfo}[2]{#2}
\providecommand{\eprint}[2][]{\url{#2}}

\bibitem[{\citenamefont{{Castro Neto} et~al.}(2009)\citenamefont{{Castro Neto},
  Guinea, Peres, Novoselov, and Geim}}]{rmp}
\bibinfo{author}{\bibfnamefont{A.~H.} \bibnamefont{{Castro Neto}}},
  \bibinfo{author}{\bibfnamefont{F.}~\bibnamefont{Guinea}},
  \bibinfo{author}{\bibfnamefont{N.~M.~R.} \bibnamefont{Peres}},
  \bibinfo{author}{\bibfnamefont{K.~S.} \bibnamefont{Novoselov}},
  \bibnamefont{and} \bibinfo{author}{\bibfnamefont{A.~K.} \bibnamefont{Geim}},
  \bibinfo{journal}{Rev. Mod. Phys.} \textbf{\bibinfo{volume}{81}},
  \bibinfo{pages}{109} (\bibinfo{year}{2009}).

\bibitem[{\citenamefont{Peres}(2010)}]{rmpPeres}
\bibinfo{author}{\bibfnamefont{N.~M.~R.} \bibnamefont{Peres}},
  \bibinfo{journal}{Rev. Mod. Phys.} \textbf{\bibinfo{volume}{82}},
  \bibinfo{pages}{2673} (\bibinfo{year}{2010}).

\bibitem[{\citenamefont{Sarma et~al.}(2011)\citenamefont{Sarma, Adam, Hwang,
  and Rossi}}]{rmpSarma}
\bibinfo{author}{\bibfnamefont{S.~D.} \bibnamefont{Sarma}},
  \bibinfo{author}{\bibfnamefont{S.}~\bibnamefont{Adam}},
  \bibinfo{author}{\bibfnamefont{E.~H.} \bibnamefont{Hwang}}, \bibnamefont{and}
  \bibinfo{author}{\bibfnamefont{E.}~\bibnamefont{Rossi}},
  \bibinfo{journal}{Rev. Mod. Phys.} \textbf{\bibinfo{volume}{83}},
  \bibinfo{pages}{407} (\bibinfo{year}{2011}).

\bibitem[{\citenamefont{Peres et~al.}(2006)\citenamefont{Peres, Guinea, and
  {Castro Neto}}}]{nmrPRB06}
\bibinfo{author}{\bibfnamefont{N.~M.~R.} \bibnamefont{Peres}},
  \bibinfo{author}{\bibfnamefont{F.}~\bibnamefont{Guinea}}, \bibnamefont{and}
  \bibinfo{author}{\bibfnamefont{A.~H.} \bibnamefont{{Castro Neto}}},
  \bibinfo{journal}{Phys. Rev. B} \textbf{\bibinfo{volume}{73}},
  \bibinfo{pages}{125411} (\bibinfo{year}{2006}).

\bibitem[{\citenamefont{Falkovsky and Pershoguba}(2007)}]{falkovsky}
\bibinfo{author}{\bibfnamefont{L.~A.} \bibnamefont{Falkovsky}}
  \bibnamefont{and} \bibinfo{author}{\bibfnamefont{S.~S.}
  \bibnamefont{Pershoguba}}, \bibinfo{journal}{Phys. Rev. B}
  \textbf{\bibinfo{volume}{76}}, \bibinfo{pages}{153410}
  (\bibinfo{year}{2007}).

\bibitem[{\citenamefont{Stauber et~al.}(2007)\citenamefont{Stauber, Peres, and
  Guinea}}]{stauberBZ}
\bibinfo{author}{\bibfnamefont{T.}~\bibnamefont{Stauber}},
  \bibinfo{author}{\bibfnamefont{N.~M.~R.} \bibnamefont{Peres}},
  \bibnamefont{and} \bibinfo{author}{\bibfnamefont{F.}~\bibnamefont{Guinea}},
  \bibinfo{journal}{Phys. Rev. B} \textbf{\bibinfo{volume}{76}},
  \bibinfo{pages}{205423} (\bibinfo{year}{2007}).

\bibitem[{\citenamefont{Stauber
  et~al.}(2008{\natexlab{a}})\citenamefont{Stauber, Peres, and {Castro
  Neto}}}]{stauberphonons}
\bibinfo{author}{\bibfnamefont{T.}~\bibnamefont{Stauber}},
  \bibinfo{author}{\bibfnamefont{N.~M.~R.} \bibnamefont{Peres}},
  \bibnamefont{and} \bibinfo{author}{\bibfnamefont{A.~H.} \bibnamefont{{Castro
  Neto}}}, \bibinfo{journal}{Phys. Rev. B} \textbf{\bibinfo{volume}{78}},
  \bibinfo{pages}{085418} (\bibinfo{year}{2008}{\natexlab{a}}).

\bibitem[{\citenamefont{Stauber
  et~al.}(2008{\natexlab{b}})\citenamefont{Stauber, Peres, and
  Geim}}]{StauberGeim}
\bibinfo{author}{\bibfnamefont{T.}~\bibnamefont{Stauber}},
  \bibinfo{author}{\bibfnamefont{N.~M.~R.} \bibnamefont{Peres}},
  \bibnamefont{and} \bibinfo{author}{\bibfnamefont{A.~K.} \bibnamefont{Geim}},
  \bibinfo{journal}{Phys. Rev. B} \textbf{\bibinfo{volume}{78}},
  \bibinfo{pages}{085432} (\bibinfo{year}{2008}{\natexlab{b}}).

\bibitem[{\citenamefont{Gusynin et~al.}(2009)\citenamefont{Gusynin, Sharapov,
  and Carbotte}}]{carbotte}
\bibinfo{author}{\bibfnamefont{V.~P.} \bibnamefont{Gusynin}},
  \bibinfo{author}{\bibfnamefont{S.~G.} \bibnamefont{Sharapov}},
  \bibnamefont{and} \bibinfo{author}{\bibfnamefont{J.~P.}
  \bibnamefont{Carbotte}}, \bibinfo{journal}{New J. Phys.}
  \textbf{\bibinfo{volume}{11}}, \bibinfo{pages}{095013}
  (\bibinfo{year}{2009}).

\bibitem[{\citenamefont{Grushin et~al.}(2009)\citenamefont{Grushin, Valenzuela,
  and Vozmediano}}]{Juan}
\bibinfo{author}{\bibfnamefont{A.~G.} \bibnamefont{Grushin}},
  \bibinfo{author}{\bibfnamefont{B.}~\bibnamefont{Valenzuela}},
  \bibnamefont{and} \bibinfo{author}{\bibfnamefont{M.~A.~H.}
  \bibnamefont{Vozmediano}}, \bibinfo{journal}{Phys. Rev. B}
  \textbf{\bibinfo{volume}{80}}, \bibinfo{pages}{155417}
  (\bibinfo{year}{2009}).

\bibitem[{\citenamefont{Mishchenko}(2009)}]{Mishchenko}
\bibinfo{author}{\bibfnamefont{E.~G.} \bibnamefont{Mishchenko}},
  \bibinfo{journal}{Phys. Rev. Lett.} \textbf{\bibinfo{volume}{103}},
  \bibinfo{pages}{246802} (\bibinfo{year}{2009}).

\bibitem[{\citenamefont{Yang et~al.}(2009)\citenamefont{Yang, Deslippe, Park,
  Cohen, and Louie}}]{LiLouie}
\bibinfo{author}{\bibfnamefont{L.}~\bibnamefont{Yang}},
  \bibinfo{author}{\bibfnamefont{J.}~\bibnamefont{Deslippe}},
  \bibinfo{author}{\bibfnamefont{C.-H.} \bibnamefont{Park}},
  \bibinfo{author}{\bibfnamefont{M.~L.} \bibnamefont{Cohen}}, \bibnamefont{and}
  \bibinfo{author}{\bibfnamefont{S.~G.} \bibnamefont{Louie}},
  \bibinfo{journal}{Phys. Rev. Lett.} \textbf{\bibinfo{volume}{103}},
  \bibinfo{pages}{186802} (\bibinfo{year}{2009}).

\bibitem[{\citenamefont{Peres et~al.}(2010)\citenamefont{Peres, Ribeiro, and
  {Castro Neto}}}]{PRL}
\bibinfo{author}{\bibfnamefont{N.~M.~R.} \bibnamefont{Peres}},
  \bibinfo{author}{\bibfnamefont{R.~M.} \bibnamefont{Ribeiro}},
  \bibnamefont{and} \bibinfo{author}{\bibfnamefont{A.~H.} \bibnamefont{{Castro
  Neto}}}, \bibinfo{journal}{Phys. Rev. Lett.} \textbf{\bibinfo{volume}{105}},
  \bibinfo{pages}{055501} (\bibinfo{year}{2010}).

\bibitem[{\citenamefont{Ferreira et~al.}(2011)\citenamefont{Ferreira,
  {Viana-Gomes}, Bludov, Pereira, Peres, and {Castro Neto}}}]{aires}
\bibinfo{author}{\bibfnamefont{A.}~\bibnamefont{Ferreira}},
  \bibinfo{author}{\bibfnamefont{J.}~\bibnamefont{{Viana-Gomes}}},
  \bibinfo{author}{\bibfnamefont{Y.~V.} \bibnamefont{Bludov}},
  \bibinfo{author}{\bibfnamefont{V.~M.} \bibnamefont{Pereira}},
  \bibinfo{author}{\bibfnamefont{N.~M.~R.} \bibnamefont{Peres}},
  \bibnamefont{and} \bibinfo{author}{\bibfnamefont{A.~H.} \bibnamefont{{Castro
  Neto}}}, \bibinfo{journal}{Phys. Rev. B} \textbf{\bibinfo{volume}{84}},
  \bibinfo{pages}{235410} (\bibinfo{year}{2011}).

\bibitem[{\citenamefont{Liu et~al.}(2012)\citenamefont{Liu, Liu, Li, Li, and
  Huang}}]{APLPhotonic}
\bibinfo{author}{\bibfnamefont{J.-T.} \bibnamefont{Liu}},
  \bibinfo{author}{\bibfnamefont{N.-H.} \bibnamefont{Liu}},
  \bibinfo{author}{\bibfnamefont{J.}~\bibnamefont{Li}},
  \bibinfo{author}{\bibfnamefont{X.~J.} \bibnamefont{Li}}, \bibnamefont{and}
  \bibinfo{author}{\bibfnamefont{J.-H.} \bibnamefont{Huang}},
  \bibinfo{journal}{Appl. Phys. Lett.} \textbf{\bibinfo{volume}{101}},
  \bibinfo{pages}{052104} (\bibinfo{year}{2012}).

\bibitem[{\citenamefont{Busl et~al.}(2012)\citenamefont{Busl, Platero, and
  Jauho}}]{Platero}
\bibinfo{author}{\bibfnamefont{M.}~\bibnamefont{Busl}},
  \bibinfo{author}{\bibfnamefont{G.}~\bibnamefont{Platero}}, \bibnamefont{and}
  \bibinfo{author}{\bibfnamefont{A.-P.} \bibnamefont{Jauho}},
  \bibinfo{journal}{Phys. Rev. B} \textbf{\bibinfo{volume}{85}},
  \bibinfo{pages}{155449} (\bibinfo{year}{2012}).

\bibitem[{\citenamefont{Nair et~al.}(2008)\citenamefont{Nair, Blake,
  Grigorenko, Novoselov, Booth, Stauber, Peres, and Geim}}]{nair}
\bibinfo{author}{\bibfnamefont{R.~R.} \bibnamefont{Nair}},
  \bibinfo{author}{\bibfnamefont{P.}~\bibnamefont{Blake}},
  \bibinfo{author}{\bibfnamefont{A.~N.} \bibnamefont{Grigorenko}},
  \bibinfo{author}{\bibfnamefont{K.~S.} \bibnamefont{Novoselov}},
  \bibinfo{author}{\bibfnamefont{T.~J.} \bibnamefont{Booth}},
  \bibinfo{author}{\bibfnamefont{T.}~\bibnamefont{Stauber}},
  \bibinfo{author}{\bibfnamefont{N.~M.~R.} \bibnamefont{Peres}},
  \bibnamefont{and} \bibinfo{author}{\bibfnamefont{A.}~\bibnamefont{Geim}},
  \bibinfo{journal}{Science} \textbf{\bibinfo{volume}{320}},
  \bibinfo{pages}{1308} (\bibinfo{year}{2008}).

\bibitem[{\citenamefont{Kuzmenko et~al.}(2008)\citenamefont{Kuzmenko, van
  Heumen, Carbone, and {van der Marel}}}]{kuzmenko}
\bibinfo{author}{\bibfnamefont{A.~B.} \bibnamefont{Kuzmenko}},
  \bibinfo{author}{\bibfnamefont{E.}~\bibnamefont{van Heumen}},
  \bibinfo{author}{\bibfnamefont{F.}~\bibnamefont{Carbone}}, \bibnamefont{and}
  \bibinfo{author}{\bibfnamefont{D.}~\bibnamefont{{van der Marel}}},
  \bibinfo{journal}{Phys. Rev. Lett.} \textbf{\bibinfo{volume}{100}},
  \bibinfo{pages}{117401} (\bibinfo{year}{2008}).

\bibitem[{\citenamefont{Mak et~al.}(2008)\citenamefont{Mak, Sfeir, Wu, Lui,
  Misewich, and Heinz}}]{mak}
\bibinfo{author}{\bibfnamefont{K.~F.} \bibnamefont{Mak}},
  \bibinfo{author}{\bibfnamefont{M.~Y.} \bibnamefont{Sfeir}},
  \bibinfo{author}{\bibfnamefont{Y.}~\bibnamefont{Wu}},
  \bibinfo{author}{\bibfnamefont{C.~H.} \bibnamefont{Lui}},
  \bibinfo{author}{\bibfnamefont{J.~A.} \bibnamefont{Misewich}},
  \bibnamefont{and} \bibinfo{author}{\bibfnamefont{T.~F.} \bibnamefont{Heinz}},
  \bibinfo{journal}{Phys. Rev. Lett.} \textbf{\bibinfo{volume}{101}},
  \bibinfo{pages}{196405} (\bibinfo{year}{2008}).

\bibitem[{\citenamefont{Li et~al.}(2008)\citenamefont{Li, Henriksen, Jiang,
  Hao, Martin, Kim, Stormer, and Basov}}]{basov}
\bibinfo{author}{\bibfnamefont{Z.~Q.} \bibnamefont{Li}},
  \bibinfo{author}{\bibfnamefont{E.}~\bibnamefont{Henriksen}},
  \bibinfo{author}{\bibfnamefont{Z.}~\bibnamefont{Jiang}},
  \bibinfo{author}{\bibfnamefont{Z.}~\bibnamefont{Hao}},
  \bibinfo{author}{\bibfnamefont{M.~C.} \bibnamefont{Martin}},
  \bibinfo{author}{\bibfnamefont{P.}~\bibnamefont{Kim}},
  \bibinfo{author}{\bibfnamefont{H.}~\bibnamefont{Stormer}}, \bibnamefont{and}
  \bibinfo{author}{\bibfnamefont{D.}~\bibnamefont{Basov}},
  \bibinfo{journal}{Nature Phys.} \textbf{\bibinfo{volume}{4}},
  \bibinfo{pages}{532} (\bibinfo{year}{2008}).

\bibitem[{\citenamefont{Wang et~al.}(2008)\citenamefont{Wang, Zhang, Tian,
  Girit, Zettl, Crommie, and Shen}}]{Crommieopt}
\bibinfo{author}{\bibfnamefont{F.}~\bibnamefont{Wang}},
  \bibinfo{author}{\bibfnamefont{Y.}~\bibnamefont{Zhang}},
  \bibinfo{author}{\bibfnamefont{C.}~\bibnamefont{Tian}},
  \bibinfo{author}{\bibfnamefont{C.}~\bibnamefont{Girit}},
  \bibinfo{author}{\bibfnamefont{A.}~\bibnamefont{Zettl}},
  \bibinfo{author}{\bibfnamefont{M.}~\bibnamefont{Crommie}}, \bibnamefont{and}
  \bibinfo{author}{\bibfnamefont{Y.~R.} \bibnamefont{Shen}},
  \bibinfo{journal}{Science} \textbf{\bibinfo{volume}{320}},
  \bibinfo{pages}{206} (\bibinfo{year}{2008}).

\bibitem[{\citenamefont{Kuzmenko et~al.}(2009)\citenamefont{Kuzmenko, Crassee,
  van~der Marel, Blake, and Novoselov}}]{kuzmenko2}
\bibinfo{author}{\bibfnamefont{A.~B.} \bibnamefont{Kuzmenko}},
  \bibinfo{author}{\bibfnamefont{I.}~\bibnamefont{Crassee}},
  \bibinfo{author}{\bibfnamefont{D.}~\bibnamefont{van~der Marel}},
  \bibinfo{author}{\bibfnamefont{P.}~\bibnamefont{Blake}}, \bibnamefont{and}
  \bibinfo{author}{\bibfnamefont{K.~S.} \bibnamefont{Novoselov}},
  \bibinfo{journal}{Phys. Rev. B} \textbf{\bibinfo{volume}{80}},
  \bibinfo{pages}{165406} (\bibinfo{year}{2009}).

\bibitem[{\citenamefont{Crassee et~al.}(2011)\citenamefont{Crassee, Levallois,
  Walter, Ostler, Bostwick, Rotenberg, Seyller, {van der Marel}, and
  Kuzmenko}}]{kuzmenkoFaraday}
\bibinfo{author}{\bibfnamefont{I.}~\bibnamefont{Crassee}},
  \bibinfo{author}{\bibfnamefont{J.}~\bibnamefont{Levallois}},
  \bibinfo{author}{\bibfnamefont{A.~L.} \bibnamefont{Walter}},
  \bibinfo{author}{\bibfnamefont{M.}~\bibnamefont{Ostler}},
  \bibinfo{author}{\bibfnamefont{A.}~\bibnamefont{Bostwick}},
  \bibinfo{author}{\bibfnamefont{E.}~\bibnamefont{Rotenberg}},
  \bibinfo{author}{\bibfnamefont{T.}~\bibnamefont{Seyller}},
  \bibinfo{author}{\bibfnamefont{D.}~\bibnamefont{{van der Marel}}},
  \bibnamefont{and} \bibinfo{author}{\bibfnamefont{A.~B.}
  \bibnamefont{Kuzmenko}}, \bibinfo{journal}{Nat. Phys.}
  \textbf{\bibinfo{volume}{7}}, \bibinfo{pages}{48} (\bibinfo{year}{2011}).

\bibitem[{\citenamefont{Bao et~al.}(2011)\citenamefont{Bao, Zhang, Wang, Ni,
  Lim, Wang, Tang, and Loh}}]{NatureLoh}
\bibinfo{author}{\bibfnamefont{Q.}~\bibnamefont{Bao}},
  \bibinfo{author}{\bibfnamefont{H.}~\bibnamefont{Zhang}},
  \bibinfo{author}{\bibfnamefont{B.}~\bibnamefont{Wang}},
  \bibinfo{author}{\bibfnamefont{Z.}~\bibnamefont{Ni}},
  \bibinfo{author}{\bibfnamefont{C.~H. Y.~X.} \bibnamefont{Lim}},
  \bibinfo{author}{\bibfnamefont{Y.}~\bibnamefont{Wang}},
  \bibinfo{author}{\bibfnamefont{D.~Y.} \bibnamefont{Tang}}, \bibnamefont{and}
  \bibinfo{author}{\bibfnamefont{K.~P.} \bibnamefont{Loh}},
  \bibinfo{journal}{Nature Photonics} \textbf{\bibinfo{volume}{5}},
  \bibinfo{pages}{411} (\bibinfo{year}{2011}).

\bibitem[{\citenamefont{Dawlaty et~al.}(2008)\citenamefont{Dawlaty, Shivaraman,
  Strait1, George, Chandrashekhar, Rana, Spencer, Veksler, and
  Chen}}]{MeasureTHzVis}
\bibinfo{author}{\bibfnamefont{J.~M.} \bibnamefont{Dawlaty}},
  \bibinfo{author}{\bibfnamefont{S.}~\bibnamefont{Shivaraman}},
  \bibinfo{author}{\bibfnamefont{J.}~\bibnamefont{Strait1}},
  \bibinfo{author}{\bibfnamefont{P.}~\bibnamefont{George}},
  \bibinfo{author}{\bibfnamefont{M.}~\bibnamefont{Chandrashekhar}},
  \bibinfo{author}{\bibfnamefont{F.}~\bibnamefont{Rana}},
  \bibinfo{author}{\bibfnamefont{M.~G.} \bibnamefont{Spencer}},
  \bibinfo{author}{\bibfnamefont{D.}~\bibnamefont{Veksler}}, \bibnamefont{and}
  \bibinfo{author}{\bibfnamefont{Y.}~\bibnamefont{Chen}},
  \bibinfo{journal}{Appl. Phys. Lett.} \textbf{\bibinfo{volume}{93}},
  \bibinfo{pages}{131905} (\bibinfo{year}{2008}).

\bibitem[{\citenamefont{Yan et~al.}(2011)\citenamefont{Yan, Xia, Zhu, Freitag,
  Dimitrakopoulos, Bol, Tulevski, and Avouris}}]{AvourisIR}
\bibinfo{author}{\bibfnamefont{H.}~\bibnamefont{Yan}},
  \bibinfo{author}{\bibfnamefont{F.}~\bibnamefont{Xia}},
  \bibinfo{author}{\bibfnamefont{W.}~\bibnamefont{Zhu}},
  \bibinfo{author}{\bibfnamefont{M.}~\bibnamefont{Freitag}},
  \bibinfo{author}{\bibfnamefont{C.}~\bibnamefont{Dimitrakopoulos}},
  \bibinfo{author}{\bibfnamefont{A.~A.} \bibnamefont{Bol}},
  \bibinfo{author}{\bibfnamefont{G.}~\bibnamefont{Tulevski}}, \bibnamefont{and}
  \bibinfo{author}{\bibfnamefont{P.}~\bibnamefont{Avouris}},
  \bibinfo{journal}{ACS Nano} \textbf{\bibinfo{volume}{5}},
  \bibinfo{pages}{9854} (\bibinfo{year}{2011}).

\bibitem[{\citenamefont{Ren et~al.}(2012{\natexlab{a}})\citenamefont{Ren,
  Zhang, Nanot, Kawayama, Tonouchi, and Kono}}]{Ren}
\bibinfo{author}{\bibfnamefont{L.}~\bibnamefont{Ren}},
  \bibinfo{author}{\bibfnamefont{Q.}~\bibnamefont{Zhang}},
  \bibinfo{author}{\bibfnamefont{S.}~\bibnamefont{Nanot}},
  \bibinfo{author}{\bibfnamefont{I.}~\bibnamefont{Kawayama}},
  \bibinfo{author}{\bibfnamefont{M.}~\bibnamefont{Tonouchi}}, \bibnamefont{and}
  \bibinfo{author}{\bibfnamefont{J.}~\bibnamefont{Kono}},
  \bibinfo{journal}{Journal of Infrared, Millimeter, and Terahertz Waves}
  \textbf{\bibinfo{volume}{33}}, \bibinfo{pages}{846}
  (\bibinfo{year}{2012}{\natexlab{a}}).

\bibitem[{\citenamefont{Ren et~al.}(2012{\natexlab{b}})\citenamefont{Ren,
  Zhang, Yao, Sun, Kaneko, Yan, Nanot, Jin, Kawayama, Tonouchi et~al.}}]{Ren2}
\bibinfo{author}{\bibfnamefont{L.}~\bibnamefont{Ren}},
  \bibinfo{author}{\bibfnamefont{Q.}~\bibnamefont{Zhang}},
  \bibinfo{author}{\bibfnamefont{J.}~\bibnamefont{Yao}},
  \bibinfo{author}{\bibfnamefont{Z.}~\bibnamefont{Sun}},
  \bibinfo{author}{\bibfnamefont{R.}~\bibnamefont{Kaneko}},
  \bibinfo{author}{\bibfnamefont{Z.}~\bibnamefont{Yan}},
  \bibinfo{author}{\bibfnamefont{S.~L.} \bibnamefont{Nanot}},
  \bibinfo{author}{\bibfnamefont{Z.}~\bibnamefont{Jin}},
  \bibinfo{author}{\bibfnamefont{I.}~\bibnamefont{Kawayama}},
  \bibinfo{author}{\bibfnamefont{M.}~\bibnamefont{Tonouchi}},
  \bibnamefont{et~al.}, \bibinfo{journal}{Nano Lett.}
  \textbf{\bibinfo{volume}{12}}, \bibinfo{pages}{3711}
  (\bibinfo{year}{2012}{\natexlab{b}}).

\bibitem[{\citenamefont{Tonouchi}(2007)}]{THz}
\bibinfo{author}{\bibfnamefont{M.}~\bibnamefont{Tonouchi}},
  \bibinfo{journal}{Nature Photonics} \textbf{\bibinfo{volume}{1}},
  \bibinfo{pages}{97} (\bibinfo{year}{2007}).

\bibitem[{\citenamefont{Yan et~al.}(2012)\citenamefont{Yan, Li, Chandra,
  Tulevski, Wu, Freitag, Zhu, Avouris, and Xia}}]{avouris}
\bibinfo{author}{\bibfnamefont{H.}~\bibnamefont{Yan}},
  \bibinfo{author}{\bibfnamefont{X.}~\bibnamefont{Li}},
  \bibinfo{author}{\bibfnamefont{B.}~\bibnamefont{Chandra}},
  \bibinfo{author}{\bibfnamefont{G.}~\bibnamefont{Tulevski}},
  \bibinfo{author}{\bibfnamefont{Y.}~\bibnamefont{Wu}},
  \bibinfo{author}{\bibfnamefont{M.}~\bibnamefont{Freitag}},
  \bibinfo{author}{\bibfnamefont{W.}~\bibnamefont{Zhu}},
  \bibinfo{author}{\bibfnamefont{P.}~\bibnamefont{Avouris}}, \bibnamefont{and}
  \bibinfo{author}{\bibfnamefont{F.}~\bibnamefont{Xia}},
  \bibinfo{journal}{Nature Nano.} \textbf{\bibinfo{volume}{7}},
  \bibinfo{pages}{330} (\bibinfo{year}{2012}).

\bibitem[{\citenamefont{Chen et~al.}(2012)\citenamefont{Chen, Badioli,
  {Alonso-Gonzalez}, Thongrattanasiri, Huth, Osmond, Spasenovic, Centeno,
  Pesquera, Godignon et~al.}}]{koppens}
\bibinfo{author}{\bibfnamefont{J.}~\bibnamefont{Chen}},
  \bibinfo{author}{\bibfnamefont{M.}~\bibnamefont{Badioli}},
  \bibinfo{author}{\bibfnamefont{P.}~\bibnamefont{{Alonso-Gonzalez}}},
  \bibinfo{author}{\bibfnamefont{S.}~\bibnamefont{Thongrattanasiri}},
  \bibinfo{author}{\bibfnamefont{F.}~\bibnamefont{Huth}},
  \bibinfo{author}{\bibfnamefont{J.}~\bibnamefont{Osmond}},
  \bibinfo{author}{\bibfnamefont{M.}~\bibnamefont{Spasenovic}},
  \bibinfo{author}{\bibfnamefont{A.}~\bibnamefont{Centeno}},
  \bibinfo{author}{\bibfnamefont{A.}~\bibnamefont{Pesquera}},
  \bibinfo{author}{\bibfnamefont{P.}~\bibnamefont{Godignon}},
  \bibnamefont{et~al.}, \bibinfo{journal}{Nature}
  \textbf{\bibinfo{volume}{487}}, \bibinfo{pages}{77} (\bibinfo{year}{2012}).

\bibitem[{\citenamefont{Konstantatos et~al.}(2012)\citenamefont{Konstantatos,
  Badioli, Gaudreau, Osmond, Bernechea, de~Arquer, Gatti, and
  Koppens}}]{koppensphoto}
\bibinfo{author}{\bibfnamefont{G.}~\bibnamefont{Konstantatos}},
  \bibinfo{author}{\bibfnamefont{M.}~\bibnamefont{Badioli}},
  \bibinfo{author}{\bibfnamefont{L.}~\bibnamefont{Gaudreau}},
  \bibinfo{author}{\bibfnamefont{J.}~\bibnamefont{Osmond}},
  \bibinfo{author}{\bibfnamefont{M.}~\bibnamefont{Bernechea}},
  \bibinfo{author}{\bibfnamefont{P.~G.} \bibnamefont{de~Arquer}},
  \bibinfo{author}{\bibfnamefont{F.}~\bibnamefont{Gatti}}, \bibnamefont{and}
  \bibinfo{author}{\bibfnamefont{F.~H.~L.} \bibnamefont{Koppens}},
  \bibinfo{journal}{Nature Nanotechnology} \textbf{\bibinfo{volume}{7}},
  \bibinfo{pages}{363} (\bibinfo{year}{2012}).

\bibitem[{\citenamefont{Fei et~al.}(2011)\citenamefont{Fei, Andreev, Bao,
  Zhang, McLeod, Wang, Stewart, Zhao, Dominguez, Thiemens
  et~al.}}]{BasovPlasmonsI}
\bibinfo{author}{\bibfnamefont{Z.}~\bibnamefont{Fei}},
  \bibinfo{author}{\bibfnamefont{G.~O.} \bibnamefont{Andreev}},
  \bibinfo{author}{\bibfnamefont{W.}~\bibnamefont{Bao}},
  \bibinfo{author}{\bibfnamefont{L.~M.} \bibnamefont{Zhang}},
  \bibinfo{author}{\bibfnamefont{A.~S.} \bibnamefont{McLeod}},
  \bibinfo{author}{\bibfnamefont{C.}~\bibnamefont{Wang}},
  \bibinfo{author}{\bibfnamefont{M.~K.} \bibnamefont{Stewart}},
  \bibinfo{author}{\bibfnamefont{Z.}~\bibnamefont{Zhao}},
  \bibinfo{author}{\bibfnamefont{G.}~\bibnamefont{Dominguez}},
  \bibinfo{author}{\bibfnamefont{M.}~\bibnamefont{Thiemens}},
  \bibnamefont{et~al.}, \bibinfo{journal}{Nano Lett.}
  \textbf{\bibinfo{volume}{11}}, \bibinfo{pages}{4701} (\bibinfo{year}{2011}).

\bibitem[{\citenamefont{Fei et~al.}(2012)\citenamefont{Fei, Rodin, Andreev,
  Bao, McLeod, Wagner, Zhang, Zhao, Dominguez, Thiemens
  et~al.}}]{BasovPlasmonsII}
\bibinfo{author}{\bibfnamefont{Z.}~\bibnamefont{Fei}},
  \bibinfo{author}{\bibfnamefont{A.~S.} \bibnamefont{Rodin}},
  \bibinfo{author}{\bibfnamefont{G.~O.} \bibnamefont{Andreev}},
  \bibinfo{author}{\bibfnamefont{W.}~\bibnamefont{Bao}},
  \bibinfo{author}{\bibfnamefont{A.~S.} \bibnamefont{McLeod}},
  \bibinfo{author}{\bibfnamefont{M.}~\bibnamefont{Wagner}},
  \bibinfo{author}{\bibfnamefont{L.~M.} \bibnamefont{Zhang}},
  \bibinfo{author}{\bibfnamefont{Z.}~\bibnamefont{Zhao}},
  \bibinfo{author}{\bibfnamefont{G.}~\bibnamefont{Dominguez}},
  \bibinfo{author}{\bibfnamefont{M.}~\bibnamefont{Thiemens}},
  \bibnamefont{et~al.}, \bibinfo{journal}{Nature}
  \textbf{\bibinfo{volume}{487}}, \bibinfo{pages}{82} (\bibinfo{year}{2012}).

\bibitem[{\citenamefont{Vicarelli et~al.}(2012)\citenamefont{Vicarelli,
  Vitiello, Coquillat, Lombardo, Ferrari, Knap, Polini, Pellegrini, and
  Tredicucci}}]{Pellegrini}
\bibinfo{author}{\bibfnamefont{L.}~\bibnamefont{Vicarelli}},
  \bibinfo{author}{\bibfnamefont{M.~S.} \bibnamefont{Vitiello}},
  \bibinfo{author}{\bibfnamefont{D.}~\bibnamefont{Coquillat}},
  \bibinfo{author}{\bibfnamefont{A.}~\bibnamefont{Lombardo}},
  \bibinfo{author}{\bibfnamefont{A.~C.} \bibnamefont{Ferrari}},
  \bibinfo{author}{\bibfnamefont{W.}~\bibnamefont{Knap}},
  \bibinfo{author}{\bibfnamefont{M.}~\bibnamefont{Polini}},
  \bibinfo{author}{\bibfnamefont{V.}~\bibnamefont{Pellegrini}},
  \bibnamefont{and}
  \bibinfo{author}{\bibfnamefont{A.}~\bibnamefont{Tredicucci}},
  \bibinfo{journal}{Nature Materials} \textbf{\bibinfo{volume}{11}},
  \bibinfo{pages}{865} (\bibinfo{year}{2012}).

\bibitem[{\citenamefont{Crassee et~al.}(2012)\citenamefont{Crassee, Orlita,
  Potemski, Walter, Ostler, Seyller, Gaponenko, Chen, and
  Kuzmenko}}]{KuzmenkoPlasmons}
\bibinfo{author}{\bibfnamefont{I.}~\bibnamefont{Crassee}},
  \bibinfo{author}{\bibfnamefont{M.}~\bibnamefont{Orlita}},
  \bibinfo{author}{\bibfnamefont{M.}~\bibnamefont{Potemski}},
  \bibinfo{author}{\bibfnamefont{A.~L.} \bibnamefont{Walter}},
  \bibinfo{author}{\bibfnamefont{M.}~\bibnamefont{Ostler}},
  \bibinfo{author}{\bibfnamefont{T.}~\bibnamefont{Seyller}},
  \bibinfo{author}{\bibfnamefont{I.}~\bibnamefont{Gaponenko}},
  \bibinfo{author}{\bibfnamefont{J.}~\bibnamefont{Chen}}, \bibnamefont{and}
  \bibinfo{author}{\bibfnamefont{A.~B.} \bibnamefont{Kuzmenko}},
  \bibinfo{journal}{Nano Lett.} \textbf{\bibinfo{volume}{12}},
  \bibinfo{pages}{2470} (\bibinfo{year}{2012}).

\bibitem[{\citenamefont{Echtermeyer et~al.}(2011)\citenamefont{Echtermeyer,
  Britnell, Jasnos, Lombardo, Gorbachev, Grigorenko, Geim, Ferrari, and
  Novoselov}}]{Echtermeyer}
\bibinfo{author}{\bibfnamefont{T.~J.} \bibnamefont{Echtermeyer}},
  \bibinfo{author}{\bibfnamefont{L.}~\bibnamefont{Britnell}},
  \bibinfo{author}{\bibfnamefont{P.~K.} \bibnamefont{Jasnos}},
  \bibinfo{author}{\bibfnamefont{A.}~\bibnamefont{Lombardo}},
  \bibinfo{author}{\bibfnamefont{R.~V.} \bibnamefont{Gorbachev}},
  \bibinfo{author}{\bibfnamefont{A.~N.} \bibnamefont{Grigorenko}},
  \bibinfo{author}{\bibfnamefont{A.~K.} \bibnamefont{Geim}},
  \bibinfo{author}{\bibfnamefont{A.~C.} \bibnamefont{Ferrari}},
  \bibnamefont{and} \bibinfo{author}{\bibfnamefont{K.~S.}
  \bibnamefont{Novoselov}}, \bibinfo{journal}{Nature Communications}
  \textbf{\bibinfo{volume}{2}}, \bibinfo{pages}{458} (\bibinfo{year}{2011}).

\bibitem[{\citenamefont{Peres et~al.}(2012)\citenamefont{Peres, Ferreira,
  Bludov, and Vasilevskiy}}]{nunoSPP}
\bibinfo{author}{\bibfnamefont{N.~M.~R.} \bibnamefont{Peres}},
  \bibinfo{author}{\bibfnamefont{A.}~\bibnamefont{Ferreira}},
  \bibinfo{author}{\bibfnamefont{Y.~V.} \bibnamefont{Bludov}},
  \bibnamefont{and} \bibinfo{author}{\bibfnamefont{M.~I.}
  \bibnamefont{Vasilevskiy}}, \bibinfo{journal}{J. Phys.: Condens. Matter}
  \textbf{\bibinfo{volume}{24}}, \bibinfo{pages}{245303}
  (\bibinfo{year}{2012}).

\bibitem[{\citenamefont{Bludov et~al.}(2012{\natexlab{a}})\citenamefont{Bludov,
  Peres, and Vasilevskiy}}]{YuliyPRB}
\bibinfo{author}{\bibfnamefont{Y.~V.} \bibnamefont{Bludov}},
  \bibinfo{author}{\bibfnamefont{N.~M.~R.} \bibnamefont{Peres}},
  \bibnamefont{and} \bibinfo{author}{\bibfnamefont{M.~I.}
  \bibnamefont{Vasilevskiy}}, \bibinfo{journal}{Phys. Rev. B}
  \textbf{\bibinfo{volume}{85}}, \bibinfo{pages}{245409}
  (\bibinfo{year}{2012}{\natexlab{a}}).

\bibitem[{\citenamefont{Ferreira and Peres}(2012)}]{AiresPeres}
\bibinfo{author}{\bibfnamefont{A.}~\bibnamefont{Ferreira}} \bibnamefont{and}
  \bibinfo{author}{\bibfnamefont{N.~M.~R.} \bibnamefont{Peres}},
  \bibinfo{journal}{Phys. Rev. B} \textbf{\bibinfo{volume}{86}},
  \bibinfo{pages}{205401} (\bibinfo{year}{2012}).

\bibitem[{\citenamefont{Gao et~al.}(2012)\citenamefont{Gao, Shu, Qiu, and
  Xu}}]{Gao}
\bibinfo{author}{\bibfnamefont{W.}~\bibnamefont{Gao}},
  \bibinfo{author}{\bibfnamefont{J.}~\bibnamefont{Shu}},
  \bibinfo{author}{\bibfnamefont{C.}~\bibnamefont{Qiu}}, \bibnamefont{and}
  \bibinfo{author}{\bibfnamefont{Q.}~\bibnamefont{Xu}}, \bibinfo{journal}{ACS
  Nano} \textbf{\bibinfo{volume}{6}}, \bibinfo{pages}{7806}
  (\bibinfo{year}{2012}).

\bibitem[{\citenamefont{Ferreira et~al.}(2012)\citenamefont{Ferreira, Peres,
  Ribeiro, and Stauber}}]{PeresCavity}
\bibinfo{author}{\bibfnamefont{A.}~\bibnamefont{Ferreira}},
  \bibinfo{author}{\bibfnamefont{N.~M.~R.} \bibnamefont{Peres}},
  \bibinfo{author}{\bibfnamefont{R.~M.} \bibnamefont{Ribeiro}},
  \bibnamefont{and} \bibinfo{author}{\bibfnamefont{T.}~\bibnamefont{Stauber}},
  \bibinfo{journal}{Phys. Rev. B} \textbf{\bibinfo{volume}{85}},
  \bibinfo{pages}{115438} (\bibinfo{year}{2012}).

\bibitem[{\citenamefont{Furchi et~al.}(2012)\citenamefont{Furchi, Urich,
  Pospischil, Lilley, Unterrainer, Detz, Klang, Andrews, Schrenk, Strasser
  et~al.}}]{MuellerCavity}
\bibinfo{author}{\bibfnamefont{M.}~\bibnamefont{Furchi}},
  \bibinfo{author}{\bibfnamefont{A.}~\bibnamefont{Urich}},
  \bibinfo{author}{\bibfnamefont{A.}~\bibnamefont{Pospischil}},
  \bibinfo{author}{\bibfnamefont{G.}~\bibnamefont{Lilley}},
  \bibinfo{author}{\bibfnamefont{K.}~\bibnamefont{Unterrainer}},
  \bibinfo{author}{\bibfnamefont{H.}~\bibnamefont{Detz}},
  \bibinfo{author}{\bibfnamefont{P.}~\bibnamefont{Klang}},
  \bibinfo{author}{\bibfnamefont{A.~M.} \bibnamefont{Andrews}},
  \bibinfo{author}{\bibfnamefont{W.}~\bibnamefont{Schrenk}},
  \bibinfo{author}{\bibfnamefont{G.}~\bibnamefont{Strasser}},
  \bibnamefont{et~al.}, \bibinfo{journal}{Nano Letters}
  \textbf{\bibinfo{volume}{12}}, \bibinfo{pages}{2773} (\bibinfo{year}{2012}).

\bibitem[{\citenamefont{Dubinov et~al.}(2011)\citenamefont{Dubinov, Aleshkin,
  Mitin, Otsuji, and Ryzhii}}]{Dubinov}
\bibinfo{author}{\bibfnamefont{A.~A.} \bibnamefont{Dubinov}},
  \bibinfo{author}{\bibfnamefont{V.~Y.} \bibnamefont{Aleshkin}},
  \bibinfo{author}{\bibfnamefont{V.}~\bibnamefont{Mitin}},
  \bibinfo{author}{\bibfnamefont{T.}~\bibnamefont{Otsuji}}, \bibnamefont{and}
  \bibinfo{author}{\bibfnamefont{V.}~\bibnamefont{Ryzhii}},
  \bibinfo{journal}{J. Phys.: Condens. Matter} \textbf{\bibinfo{volume}{23}},
  \bibinfo{pages}{145302} (\bibinfo{year}{2011}).

\bibitem[{\citenamefont{Zhang et~al.}(2012)\citenamefont{Zhang, Zhang, and
  Xu}}]{Xu}
\bibinfo{author}{\bibfnamefont{J.}~\bibnamefont{Zhang}},
  \bibinfo{author}{\bibfnamefont{L.}~\bibnamefont{Zhang}}, \bibnamefont{and}
  \bibinfo{author}{\bibfnamefont{W.}~\bibnamefont{Xu}}, \bibinfo{journal}{J.
  Phys. D: Appl. Phys.} \textbf{\bibinfo{volume}{45}}, \bibinfo{pages}{113001}
  (\bibinfo{year}{2012}).

\bibitem[{\citenamefont{Grigorenko et~al.}(2012)\citenamefont{Grigorenko,
  Polini, and Novoselov}}]{Kostia}
\bibinfo{author}{\bibfnamefont{A.~N.} \bibnamefont{Grigorenko}},
  \bibinfo{author}{\bibfnamefont{M.}~\bibnamefont{Polini}}, \bibnamefont{and}
  \bibinfo{author}{\bibfnamefont{K.~S.} \bibnamefont{Novoselov}},
  \bibinfo{journal}{Nature Photonics} \textbf{\bibinfo{volume}{6}},
  \bibinfo{pages}{749} (\bibinfo{year}{2012}).

\bibitem[{\citenamefont{Tassin et~al.}(2012)\citenamefont{Tassin, Koschny,
  Kafesaki, and Soukoulis}}]{Soukoulis}
\bibinfo{author}{\bibfnamefont{P.}~\bibnamefont{Tassin}},
  \bibinfo{author}{\bibfnamefont{T.}~\bibnamefont{Koschny}},
  \bibinfo{author}{\bibfnamefont{M.}~\bibnamefont{Kafesaki}}, \bibnamefont{and}
  \bibinfo{author}{\bibfnamefont{C.~M.} \bibnamefont{Soukoulis}},
  \bibinfo{journal}{Nature Photonics} \textbf{\bibinfo{volume}{6}},
  \bibinfo{pages}{259} (\bibinfo{year}{2012}).

\bibitem[{\citenamefont{Bonaccorso et~al.}(2010)\citenamefont{Bonaccorso, Sun,
  Hasan, and Ferrari}}]{Ferrari}
\bibinfo{author}{\bibfnamefont{F.}~\bibnamefont{Bonaccorso}},
  \bibinfo{author}{\bibfnamefont{Z.}~\bibnamefont{Sun}},
  \bibinfo{author}{\bibfnamefont{T.}~\bibnamefont{Hasan}}, \bibnamefont{and}
  \bibinfo{author}{\bibfnamefont{A.~C.} \bibnamefont{Ferrari}},
  \bibinfo{journal}{Nature Photonics} \textbf{\bibinfo{volume}{4}},
  \bibinfo{pages}{611} (\bibinfo{year}{2010}).

\bibitem[{\citenamefont{Bao and Loh}(2012)}]{Loh}
\bibinfo{author}{\bibfnamefont{Q.}~\bibnamefont{Bao}} \bibnamefont{and}
  \bibinfo{author}{\bibfnamefont{K.~P.} \bibnamefont{Loh}},
  \bibinfo{journal}{ACS Nano} \textbf{\bibinfo{volume}{6}},
  \bibinfo{pages}{3677} (\bibinfo{year}{2012}).

\bibitem[{\citenamefont{{Sensale-Rodriguez}
  et~al.}(2012)\citenamefont{{Sensale-Rodriguez}, Yan, Rafique, Zhu, Li, Liang,
  Gundlach, Protasenko, Kelly, Jena et~al.}}]{Modulators}
\bibinfo{author}{\bibfnamefont{B.}~\bibnamefont{{Sensale-Rodriguez}}},
  \bibinfo{author}{\bibfnamefont{R.}~\bibnamefont{Yan}},
  \bibinfo{author}{\bibfnamefont{S.}~\bibnamefont{Rafique}},
  \bibinfo{author}{\bibfnamefont{M.}~\bibnamefont{Zhu}},
  \bibinfo{author}{\bibfnamefont{W.}~\bibnamefont{Li}},
  \bibinfo{author}{\bibfnamefont{X.}~\bibnamefont{Liang}},
  \bibinfo{author}{\bibfnamefont{D.}~\bibnamefont{Gundlach}},
  \bibinfo{author}{\bibfnamefont{V.}~\bibnamefont{Protasenko}},
  \bibinfo{author}{\bibfnamefont{M.~M.} \bibnamefont{Kelly}},
  \bibinfo{author}{\bibfnamefont{D.}~\bibnamefont{Jena}}, \bibnamefont{et~al.},
  \bibinfo{journal}{Nano Lett.} \textbf{\bibinfo{volume}{12}},
  \bibinfo{pages}{4518} (\bibinfo{year}{2012}).

\bibitem[{\citenamefont{Bludov et~al.}(2010)\citenamefont{Bludov, Vasilevskiy,
  and Peres}}]{YuliyEPL}
\bibinfo{author}{\bibfnamefont{Y.~V.} \bibnamefont{Bludov}},
  \bibinfo{author}{\bibfnamefont{M.~I.} \bibnamefont{Vasilevskiy}},
  \bibnamefont{and} \bibinfo{author}{\bibfnamefont{N.~M.~R.}
  \bibnamefont{Peres}}, \bibinfo{journal}{EuroPhys. Lett.}
  \textbf{\bibinfo{volume}{92}}, \bibinfo{pages}{68001} (\bibinfo{year}{2010}).

\bibitem[{\citenamefont{Bludov et~al.}(2012{\natexlab{b}})\citenamefont{Bludov,
  Vasilevskiy, and Peres}}]{YuliyPolarizer}
\bibinfo{author}{\bibfnamefont{Y.~V.} \bibnamefont{Bludov}},
  \bibinfo{author}{\bibfnamefont{M.~I.} \bibnamefont{Vasilevskiy}},
  \bibnamefont{and} \bibinfo{author}{\bibfnamefont{N.~M.~R.}
  \bibnamefont{Peres}}, \bibinfo{journal}{J. Appl. Phys.}
  \textbf{\bibinfo{volume}{112}}, \bibinfo{pages}{084320}
  (\bibinfo{year}{2012}{\natexlab{b}}).

\bibitem[{\citenamefont{Zhou and Wu}(2011)}]{Wu}
\bibinfo{author}{\bibfnamefont{Y.}~\bibnamefont{Zhou}} \bibnamefont{and}
  \bibinfo{author}{\bibfnamefont{M.~W.} \bibnamefont{Wu}},
  \bibinfo{journal}{Phys. Rev. B} \textbf{\bibinfo{volume}{83}},
  \bibinfo{pages}{245436} (\bibinfo{year}{2011}).

\bibitem[{\citenamefont{Mikhailov}(2007)}]{Mikhailov1}
\bibinfo{author}{\bibfnamefont{S.~A.} \bibnamefont{Mikhailov}},
  \bibinfo{journal}{Europhys. Lett.} \textbf{\bibinfo{volume}{79}},
  \bibinfo{pages}{27002} (\bibinfo{year}{2007}).

\bibitem[{\citenamefont{Mikhailov and Ziegler}(2008)}]{Mikhailov2}
\bibinfo{author}{\bibfnamefont{S.~A.} \bibnamefont{Mikhailov}}
  \bibnamefont{and} \bibinfo{author}{\bibfnamefont{K.}~\bibnamefont{Ziegler}},
  \bibinfo{journal}{J. Phys.: Condens. Matter} \textbf{\bibinfo{volume}{20}},
  \bibinfo{pages}{384204} (\bibinfo{year}{2008}).

\bibitem[{\citenamefont{Mikhailov}(2009)}]{Mikhailov3}
\bibinfo{author}{\bibfnamefont{S.~A.} \bibnamefont{Mikhailov}},
  \bibinfo{journal}{Microelectronics Journal} \textbf{\bibinfo{volume}{40}},
  \bibinfo{pages}{712} (\bibinfo{year}{2009}).

\bibitem[{\citenamefont{Falkovsky}(2008)}]{Falkovsky2}
\bibinfo{author}{\bibfnamefont{L.~A.} \bibnamefont{Falkovsky}},
  \bibinfo{journal}{J. Phys.: Conf. Ser.} \textbf{\bibinfo{volume}{129}},
  \bibinfo{pages}{012004} (\bibinfo{year}{2008}).

\bibitem[{\citenamefont{Horng et~al.}(2011)\citenamefont{Horng, Chen, Geng,
  Girit, Zhang, Hao, Bechtel, Martin, Zettl, Crommie et~al.}}]{FengWang}
\bibinfo{author}{\bibfnamefont{J.}~\bibnamefont{Horng}},
  \bibinfo{author}{\bibfnamefont{C.-F.} \bibnamefont{Chen}},
  \bibinfo{author}{\bibfnamefont{B.}~\bibnamefont{Geng}},
  \bibinfo{author}{\bibfnamefont{C.}~\bibnamefont{Girit}},
  \bibinfo{author}{\bibfnamefont{Y.}~\bibnamefont{Zhang}},
  \bibinfo{author}{\bibfnamefont{Z.}~\bibnamefont{Hao}},
  \bibinfo{author}{\bibfnamefont{H.~A.} \bibnamefont{Bechtel}},
  \bibinfo{author}{\bibfnamefont{M.}~\bibnamefont{Martin}},
  \bibinfo{author}{\bibfnamefont{A.}~\bibnamefont{Zettl}},
  \bibinfo{author}{\bibfnamefont{M.~F.} \bibnamefont{Crommie}},
  \bibnamefont{et~al.}, \bibinfo{journal}{Phys. Rev. B}
  \textbf{\bibinfo{volume}{83}}, \bibinfo{pages}{165113}
  (\bibinfo{year}{2011}).

\bibitem[{\citenamefont{Chandezon et~al.}(1980)\citenamefont{Chandezon,
  Maystre, and Raoult}}]{Chandezon}
\bibinfo{author}{\bibfnamefont{J.}~\bibnamefont{Chandezon}},
  \bibinfo{author}{\bibfnamefont{D.}~\bibnamefont{Maystre}}, \bibnamefont{and}
  \bibinfo{author}{\bibfnamefont{G.}~\bibnamefont{Raoult}},
  \bibinfo{journal}{J. Optics} \textbf{\bibinfo{volume}{11}},
  \bibinfo{pages}{235} (\bibinfo{year}{1980}).

\bibitem[{\citenamefont{Li et~al.}(1999)\citenamefont{Li, Chandezon, Granet,
  and Plumey}}]{Lifeng}
\bibinfo{author}{\bibfnamefont{L.}~\bibnamefont{Li}},
  \bibinfo{author}{\bibfnamefont{J.}~\bibnamefont{Chandezon}},
  \bibinfo{author}{\bibfnamefont{G.}~\bibnamefont{Granet}}, \bibnamefont{and}
  \bibinfo{author}{\bibfnamefont{J.-P.} \bibnamefont{Plumey}},
  \bibinfo{journal}{Applied Optics} \textbf{\bibinfo{volume}{38}},
  \bibinfo{pages}{304} (\bibinfo{year}{1999}).

\bibitem[{\citenamefont{Sheng et~al.}(1982)\citenamefont{Sheng, Stepleman, and
  Sanda}}]{Ping}
\bibinfo{author}{\bibfnamefont{P.}~\bibnamefont{Sheng}},
  \bibinfo{author}{\bibfnamefont{R.~S.} \bibnamefont{Stepleman}},
  \bibnamefont{and} \bibinfo{author}{\bibfnamefont{P.~N.} \bibnamefont{Sanda}},
  \bibinfo{journal}{Phys. Rev. B} \textbf{\bibinfo{volume}{26}},
  \bibinfo{pages}{2907} (\bibinfo{year}{1982}).

\bibitem[{\citenamefont{Tongay et~al.}(2011)\citenamefont{Tongay, Berke,
  Lemaitre, Nasrollahi, Tanner, Hebard1, and Appleton}}]{chemicaldopeII}
\bibinfo{author}{\bibfnamefont{S.}~\bibnamefont{Tongay}},
  \bibinfo{author}{\bibfnamefont{K.}~\bibnamefont{Berke}},
  \bibinfo{author}{\bibfnamefont{M.}~\bibnamefont{Lemaitre}},
  \bibinfo{author}{\bibfnamefont{Z.}~\bibnamefont{Nasrollahi}},
  \bibinfo{author}{\bibfnamefont{D.~B.} \bibnamefont{Tanner}},
  \bibinfo{author}{\bibfnamefont{A.~F.} \bibnamefont{Hebard1}},
  \bibnamefont{and} \bibinfo{author}{\bibfnamefont{B.~R.}
  \bibnamefont{Appleton}}, \bibinfo{journal}{Nanotechnology}
  \textbf{\bibinfo{volume}{22}}, \bibinfo{pages}{425701}
  (\bibinfo{year}{2011}).

\bibitem[{\citenamefont{Liu et~al.}(2011)\citenamefont{Liu, Liu, and
  Zhu}}]{chemicaldopeI}
\bibinfo{author}{\bibfnamefont{H.}~\bibnamefont{Liu}},
  \bibinfo{author}{\bibfnamefont{Y.}~\bibnamefont{Liu}}, \bibnamefont{and}
  \bibinfo{author}{\bibfnamefont{D.}~\bibnamefont{Zhu}}, \bibinfo{journal}{J.
  Mater. Chem.} \textbf{\bibinfo{volume}{21}}, \bibinfo{pages}{3335}
  (\bibinfo{year}{2011}).

\bibitem[{\citenamefont{Ju et~al.}(2011)\citenamefont{Ju, Geng, Horng, Girit,
  Martin, Hao, Bechtel, Liang, Zettl, Shen et~al.}}]{LongJuPlasmonics}
\bibinfo{author}{\bibfnamefont{L.}~\bibnamefont{Ju}},
  \bibinfo{author}{\bibfnamefont{B.}~\bibnamefont{Geng}},
  \bibinfo{author}{\bibfnamefont{J.}~\bibnamefont{Horng}},
  \bibinfo{author}{\bibfnamefont{C.}~\bibnamefont{Girit}},
  \bibinfo{author}{\bibfnamefont{M.~C.} \bibnamefont{Martin}},
  \bibinfo{author}{\bibfnamefont{Z.}~\bibnamefont{Hao}},
  \bibinfo{author}{\bibfnamefont{H.~A.} \bibnamefont{Bechtel}},
  \bibinfo{author}{\bibfnamefont{X.}~\bibnamefont{Liang}},
  \bibinfo{author}{\bibfnamefont{A.}~\bibnamefont{Zettl}},
  \bibinfo{author}{\bibfnamefont{Y.~R.} \bibnamefont{Shen}},
  \bibnamefont{et~al.}, \bibinfo{journal}{Nature Nanotechnology}
  \textbf{\bibinfo{volume}{6}}, \bibinfo{pages}{630} (\bibinfo{year}{2011}).

\end{thebibliography}

\end{document}